\documentclass{article}
\usepackage{template/iclr2027_conference,times}

\usepackage{amsmath,amsfonts,bm}

\def\eqref#1{equation~\ref{#1}}

\def\1{\bm{1}}

\DeclareMathAlphabet{\mathsfit}{\encodingdefault}{\sfdefault}{m}{sl}
\SetMathAlphabet{\mathsfit}{bold}{\encodingdefault}{\sfdefault}{bx}{n}

\newcommand{\DepthSurrP}{0.183}

\newcommand{\PJbbP}{0.195}
\newcommand{\PMiP}{0.230}
\newcommand{\PSrP}{0.499}

\newcommand{\AGammaZero}{+0.63}
\newcommand{\AGammaThirty}{+0.93}
\newcommand{\AGammaThirtyLo}{0.70}
\newcommand{\AGammaThirtyHi}{1.19}
\newcommand{\ADepthDelta}{+0.29}
\newcommand{\ASlopeLate}{+0.52}
\newcommand{\ASlopeLateLo}{0.31}
\newcommand{\ASlopeLateHi}{0.76}

\newcommand{\SuffAdvFull}{0.95}
\newcommand{\SuffBandLow}{0.00}
\newcommand{\SuffBandTop}{0.25}
\newcommand{\SuffBandUnion}{0.40}

\newcommand{\AlmAsr}{0.15}
\newcommand{\AlmRandMean}{0.17}

\newcommand{\GrmAsr}{0.75}
\newcommand{\GrmRandMean}{0.86}
\newcommand{\GrmRandLo}{0.75}
\newcommand{\GrmRandHi}{0.95}
\newcommand{\GrmRandBoot}{0.84}
\newcommand{\GrmRandBootHi}{0.885}

\usepackage{amsmath}
\usepackage{amssymb}
\usepackage{bm}
\usepackage{pifont}
\usepackage{dsfont}
\usepackage{booktabs}
\usepackage{graphicx}
\usepackage{booktabs}
\usepackage{tabularx}
\usepackage{hyperref}
\usepackage{url}
\usepackage{xcolor}

\newdimen\SecBeforeNeg  \SecBeforeNeg=-0.32cm
\newdimen\SecAfterNeg   \SecAfterNeg=-0.30cm
\newdimen\SubBeforeNeg  \SubBeforeNeg=-0.30cm
\newdimen\SubAfterNeg   \SubAfterNeg=-0.20cm
\newdimen\ParaBeforeNeg \ParaBeforeNeg=-0.40cm
\newdimen\ParaVspace     \ParaVspace=-0.1cm 

\title{Where Does the Audio Jailbreak Live?\\
  A Controlled Frequency-Depth Audit of AdvWave-P on Qwen2-Audio}

\author{
    \textbf{Boyuan Chen}\textsuperscript{1,2},
    \textbf{Minseok Kim}\textsuperscript{1},
    \textbf{Sohaila Abdulsattar}\textsuperscript{1},
    \textbf{Minghao Shao}\textsuperscript{1,2},\\
    \textbf{Siddharth Garg}\textsuperscript{2},
    \textbf{Ramesh Karri}\textsuperscript{2},
    \textbf{Muhammad Shafique}\textsuperscript{1},\\
    \normalfont\textsuperscript{1}eBRAIN Lab, Division of Engineering, New York University Abu Dhabi, UAE \\
    \textsuperscript{2}Tandon School of Engineering, New York University, NY, USA \\
    \texttt{\{boyuan.chen, mk7545, sm10688, shao.minghao, sg175\}@nyu.edu} \\
    \texttt{\{rkarri, muhammad.shafique\}@nyu.edu}
}

\iclrfinalcopy

\begin{document}
\maketitle
\fancyhead{}

\begin{abstract}
\vspace{-0.3cm}
We audit frequency and decoder-depth claims for AdvWave-P, an additive audio jailbreak, on Qwen2-Audio. The protocol masks frequency components of the perturbation in the short-time Fourier transform (STFT) domain and measures attack success and audio-span representations. On 520 AdvBench prompts, the primary judge labels 76.7\% of adversarial inputs as jailbreaks. A condition-blind, single-annotator validation yields a Rogan-Gladen sensitivity estimate of 0.87 for this adversarial condition (approximately 0.83--0.95 when validation-rate uncertainty is propagated); this correction is not applied to masked conditions. The apparent frequency ranking depends on the partition: energy share alone predicts the standard eight-band ranking (Spearman's $\rho=0.95$), and equal-Hz and equal-energy partitions show that masking any tested band can sharply reduce attack success. At a finer, 16-band equal-energy resolution, however, masking the narrow $7520$--$7960$ Hz band leaves ASR at 0.10; without a matched control at this resolution, this remains unresolved. Matched-energy scattered-removal tests show that contiguous removal is more damaging in lower bands, while both forms of removal approach the floor in upper bands. Global rescaling leaves ASR near baseline, but tests amplitude sensitivity rather than frequency location. In a re-optimization pilot ($n=20$), tested single-band and two-band supports reach ASRs of 0.00, 0.25, and 0.40, while random supports covering about half the STFT bins reach a mean of 0.86. A prompt- and energy-adjusted model finds that audio-span divergence is associated with band necessity, with the coefficient increasing from $+0.63$ at the projector output to $+0.93$ at layer 30 (contrast $+0.293$, 95\% CI $[0.11,0.51]$). This association is not a causal localization, and single-layer patching does not establish a causal layer. Together, the results support a partition-aware audit of frequency claims while leaving the fine-resolution top-band result and broader generality open. Code: \url{https://github.com/BoyuanChen99/Where_Does_The_Audio_Jailbreak_Live}.
\vspace{-0.3cm}
\end{abstract}

\vspace{-0.2cm}
\section{Introduction}
\vspace{\SecAfterNeg}
\label{sec:introduction}
Audio-language models (ALMs), including Qwen2-Audio and LLaMA-Omni, extend language models with speech frontends~\citep{chu2024qwen2audio,fang2025llamaomni}. A model may refuse a harmful text instruction yet comply when the same instruction is presented as adversarial audio. AdvWave and AdvWave-P demonstrate this risk on AdvBench~\citep{kang2025advwave,jin2025almguard,zou2023universal}; related vulnerabilities have also been reported for other attacks and ALMs~\citep{song2025audiojailbreak,peri2024speechguard,yang2025audioachilles}. To evaluate frequency-localization claims, an analysis must distinguish a band's location from its width, energy, and the structure of the removed signal.

\vspace{\ParaVspace}
Prior work selects frequency regions for different purposes. GRM ranks mel bands by an attack-to-utility gradient ratio~\citep{wang2026grm}. ALMGuard selects mel bins for applying a defensive universal perturbation, balancing jailbreak mitigation against speech-recognition performance~\citep{jin2025almguard}. Gupta et al. use encoder attention to identify a broad $1$--$8$\,kHz region for universal audio jailbreaks~\citep{gupta2025iambad}. These methods do not all make the same claim about which components of an attacker's perturbation are necessary. We test how apparent band importance changes with the analysis partition, perturbation energy, and removal structure, then examine its association with decoder representations.

\vspace{\ParaVspace}
We study AdvWave-P on Qwen2-Audio and use it to audit frequency-localization claims. We mask perturbation components in the STFT domain, reconstruct each waveform, and measure attack success and decoder representations. The comparisons include random contiguous regions, equal-Hz and equal-energy partitions, matched-energy scattered removal, and global rescaling. The standard eight-band ranking is strongly associated with band energy, while the finer equal-energy sweep leaves one narrow top-frequency result unresolved because it lacks a matched control. Separately, a prompt- and energy-adjusted model finds a depth-varying association between audio-span divergence and band necessity. This association does not establish a causal layer.

\vspace{\ParaVspace}
We make four contributions, summarized with their supporting results in Table~\ref{tab:summary}.
\vspace{-0.2cm}
\begin{itemize}
\item \textbf{A controlled audit of frequency rankings} (Section~\ref{sec:methodology}). The protocol measures judge-adjudicated ASR and layer-wise audio-span divergence after STFT-domain occlusion. Random-region, equal-Hz, equal-energy, and scattered-removal comparisons expose how band rankings depend on partition and removal structure; global rescaling serves only as an amplitude check.
\vspace{-0.1cm}
\item \textbf{Frequency results depend on resolution} (Section~\ref{sec:res_asr}). Energy share alone closely predicts the standard eight-band ranking (Spearman $\rho=0.95$). Equal-Hz and eight-band equal-energy partitions show substantial effects across bands, while the narrow top band in the 16-band equal-energy sweep remains unresolved without a matched control.
\vspace{-0.1cm}
\item \textbf{Removal structure and support coverage matter} (Section~\ref{sec:res_fidelity}). At low frequencies, contiguous removal is more damaging than scattered removal of the same energy; at high frequencies, both approach the floor. In a re-optimization pilot, tested single-band and two-band supports perform poorly, while some random supports covering about half the STFT bins sustain the attack. The ALMGuard and GRM audits are inconclusive at their tested scales and do not evaluate the methods' original objectives.
\vspace{-0.1cm}
\item \textbf{A depth-varying representational association} (Sections~\ref{sec:res_repr} and~\ref{sec:res_causal}). In a prompt- and energy-adjusted model, audio-span divergence is associated with band necessity, and the estimated coefficient increases with depth. The association is not causal, and single-layer patching does not identify a causal layer.
\end{itemize}

\vspace{\SecBeforeNeg}
\section{Related Work}
\vspace{\SecAfterNeg}
\label{sec:related}

\paragraph{Audio jailbreaks on ALMs.}
Qwen2-Audio and LLaMA-Omni pair a log-mel frontend with an instruction-tuned language model~\citep{chu2024qwen2audio,fang2025llamaomni}, but text-side safety does not always transfer to audio. SpeechGuard documents vulnerabilities to adversarial waveforms~\citep{peri2024speechguard}. \citet{kang2025advwave} introduce dual-phase optimization to address gradient shattering in the encoder and report over $90\%$ ASR on AdvBench~\citep{zou2023universal}. \citet{jin2025almguard} introduce the ALMGuard defense and AdvWave-P, the additive attack evaluated here. AudioJailbreak and related work broadened evaluation across attacks and ALMs~\citep{song2025audiojailbreak,yang2025audioachilles}. We ask whether apparent frequency patterns persist when bandwidth, perturbation energy, and decoder depth are tested with matched controls.

\vspace{\ParaBeforeNeg}
\paragraph{Classical audio-adversarial and psychoacoustic attacks.}
Our analysis also draws on audio-adversarial research. \citet{carlini2018audioxe} construct targeted audio attacks. \citet{qin2019imperceptible,schoenherr2019psychoacoustic} use psychoacoustic methods to reduce audibility. These studies motivate attention to spectral structure, but AdvWave-P constrains only $\|\bm\delta\|_\infty$ and has no explicit masking-threshold objective. Our findings concern how its spectrum is analyzed, not the psychoacoustic mechanism.

\vspace{\ParaBeforeNeg}
\paragraph{Frequency-domain structure and defenses.}
GRM ranks mel bands by attack-to-utility ratio and perturbs a selected subset~\citep{wang2026grm}. ALMGuard selects mel bins for its defensive universal perturbation~\citep{jin2025almguard}, while \citet{gupta2025iambad} use encoder attention to identify a broad $1$--$8$\,kHz region for universal audio jailbreaks. \citet{yang2019temporaldep} study temporal-dependency-based detection, not a band-stop defense. Our analysis tests how frequency rankings change with the partition and removal control, then examines their association with decoder representations.

\vspace{\ParaBeforeNeg}
\paragraph{Mechanistic interpretability and safety evaluation.}
We use representation-space probes and linear CKA to examine decoder states~\citep{alain2017probes,kornblith2019linear}. We also test activation patching, drawing on causal-mediation methods~\citep{vig2020causal,meng2022locating,wang2023ioi,conmy2023acdc}. Appendix~\ref{sec:app_activation} discusses the patching protocol and its limits, including the single-layer caveat described in recent interpretability guidance~\citep{heimersheim2024patching}.

\vspace{\ParaBeforeNeg}
\paragraph{Removal-based attribution and its pitfalls.}
Occlusion attributes an output change to the region that was removed, but the effect can depend on how the removal is constructed. In vision, Zeiler and Fergus~\citep{zeiler2014visualizing} measure output changes after occluding input regions, while ROAR~\citep{hooker2019roar} retrains on the modified data to separate feature information from removal effects. Sanity checks have also shown that some saliency maps can be insensitive to what a model has learned~\citep{adebayo2018sanity}; RISE~\citep{petsiuk2018rise} addresses the variability of any single mask by averaging over random masks. Covert et al.~\citep{covert2021explaining} place these methods within a broader account of removal-based explanations. We apply the same caution to spectral occlusion: our random-band, equal-energy, and scattered-removal controls test whether a measured effect depends on band location, energy, or removal structure. The limits of single-layer patching are discussed in Section~\ref{sec:res_causal}.

\vspace{\SecBeforeNeg}
\section{Methodology}
\vspace{\SecAfterNeg}
\label{sec:methodology}

We study one audio jailbreak by removing selected frequency components of its perturbation and measuring attack behavior and decoder representations. Behavioral importance measures how each removal changes attack success. Representational divergence measures how the audio-span representation changes across decoder layers.

\vspace{\SubBeforeNeg}
\subsection{Setup and Mel-Band Occlusion}
\vspace{\SubAfterNeg}
\label{sec:setup}

\begin{figure}[!t]
  \centering
  \includegraphics[width=0.99\linewidth, trim= 25 300 450 20]{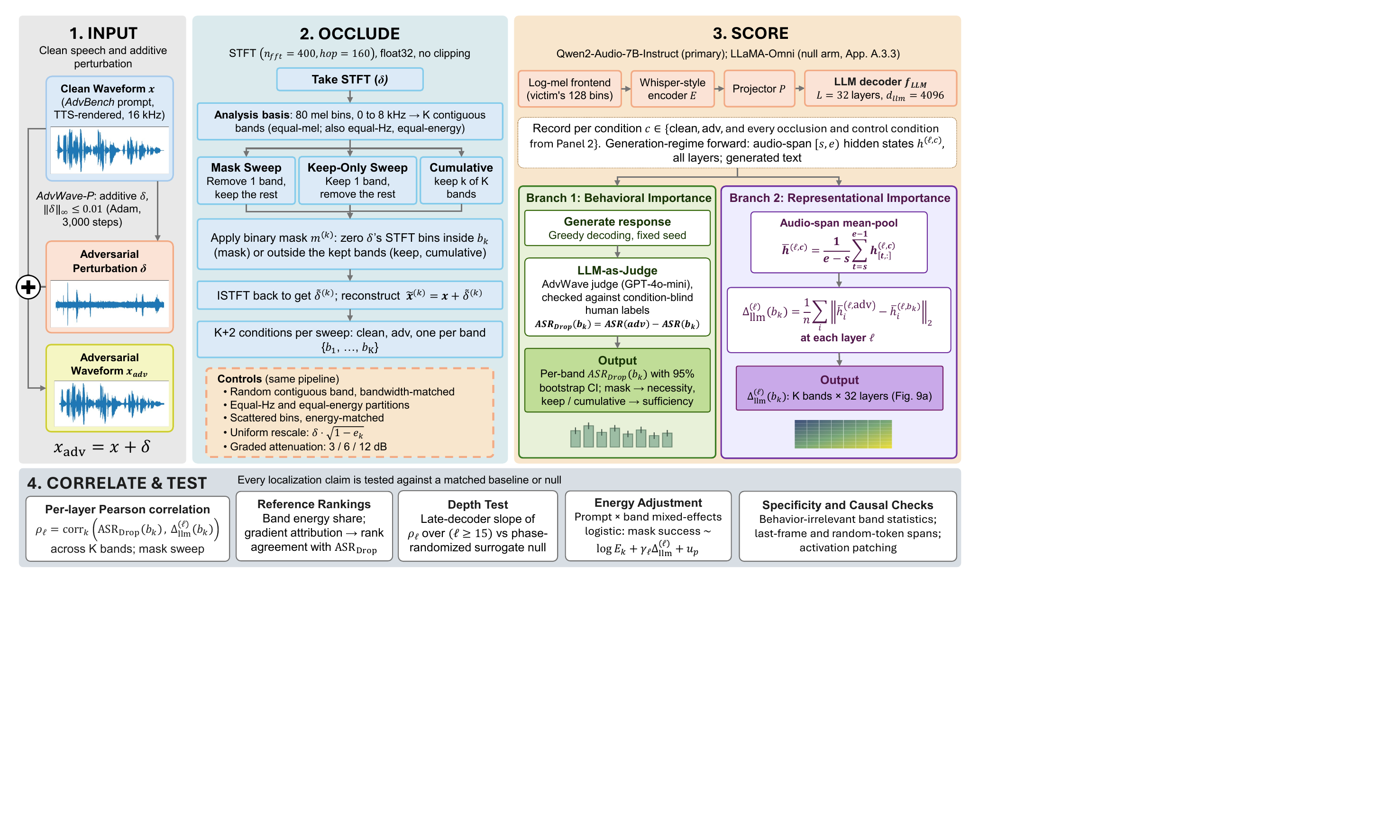}
  \caption{Protocol overview. We decompose the perturbation $\delta$ in the STFT domain, mask or retain each analysis band, reconstruct the waveform, and score the victim response. Controls include random contiguous regions, equal-Hz and equal-energy partitions, matched-energy scattered removal, global rescaling by $\sqrt{1-e_k}$, and graded attenuation. Hidden-state analysis measures layer-wise audio-span divergence $\Delta^{(\ell)}_{\mathrm{llm}}$. We compare ASR drops with band-energy and gradient rankings. A surrogate test evaluates the raw depth curve, and a prompt-by-band model estimates the energy-adjusted association. Section~\ref{sec:results} reports the findings.}
  \label{fig:methodology}
\end{figure}

Let $\mathbf{x}\in\mathbb{R}^{T}$ be a clean $16$\,kHz waveform from AdvBench~\citep{zou2023universal}, and let $\mathbf{x}_{\text{adv}}=\mathbf{x}+\bm\delta$ be its AdvWave-P counterpart~\citep{jin2025almguard}. We optimize $\bm\delta$ with $\|\bm\delta\|_\infty\le0.01$ using Adam (learning rate $10^{-3}$, $3{,}000$ iterations). We evaluate the clean and adversarial waveforms and band-occluded variants of the adversarial waveform. The victim processes each input through its own $128$-bin log-mel frontend, projects the encoded audio into tokens, and passes them to an $L$-layer language-model decoder. Our analysis uses a separate frequency basis: contiguous bands defined by mel-scale edges over $0$--$8$\,kHz. We use $K{=}8$ bands for Qwen (10 analysis bins per band), a $K{=}16$ refinement (5 bins per band), and $K{=}13$ bands on Omni's 128-bin analysis basis. This analysis basis is distinct from each victim's frontend. We report band boundaries in Hz and map them to STFT bins for reconstruction. For band $b_k$, we define the occluded waveform as
\begin{equation}
\tilde{\mathbf{x}}_{\text{mask}}^{(k)} = \mathbf{x} + \mathrm{ISTFT}\,\big(\mathrm{STFT}(\bm\delta)\odot\mathbf{m}^{(k)}\big),
\end{equation}
\vspace{\ParaVspace}
where $\mathbf{m}^{(k)}$ is $0$ on the bins covered by $b_k$ and $1$ elsewhere, zeroing the perturbation's energy inside $b_k$ while leaving the clean waveform unchanged. The \emph{mask} sweep measures necessity; the \emph{keep-only} sweep measures sufficiency, formalized in Section~\ref{sec:asr_drop}.

\vspace{-0.15cm}
\textbf{Reconstruction and reference evaluation.} All reconstructions use a single float32, no-clipping STFT-mask-ISTFT path, with exact inversion and no 16-bit quantization (Table~\ref{tab:recon_path}).

\vspace{-0.15cm}
\textbf{Protocol interface.} The protocol takes a clean waveform and a time-aligned additive perturbation. White-box access to decoder hidden states is required only for the depth analysis.

\vspace{\SubBeforeNeg}
\subsection{Behavioral Importance}
\vspace{\SubAfterNeg}
\label{sec:asr_drop}

\vspace{\ParaVspace}
We evaluate the mask and keep-only sweeps separately. Each sweep has $K{+}2$ conditions: clean, full adversarial, and one masked or retained variant per band. For each condition $c$, we generate a response using AdvWave's decoding settings and score it with the judge in Section~\ref{sec:exp_setup}. Attack success is $\mathrm{ASR}(c)=\frac{1}{n}\sum_i\mathds{1}[\text{judge}(c_i)=\text{jailbroken}]$, and band importance is $\mathrm{ASR\_Drop}(b_k)=\mathrm{ASR}(\text{adv})-\mathrm{ASR}(b_k)$. Conditions within a sweep share the same fixed seed and greedy decoding. For the representation analysis (Section~\ref{sec:repr_div}), we record hidden states and the audio-token span in a separate generation-regime forward pass using the same input and decoding configuration. We define necessity as $\mathcal{I}^{+}(b_k)=\mathrm{ASR}(\text{adv})-\mathrm{ASR}(\text{mask}(b_k))$: larger values mean that masking the band reduces ASR. Sufficiency is $\mathcal{I}^{-}(b_k)=\mathrm{ASR}(\text{adv})-\mathrm{ASR}(\text{keep}(b_k))$: smaller values mean that retaining the band better sustains the attack. We report bootstrap 95\% CIs and Spearman correlations between rankings (Section~\ref{sec:meth_inference}).

\vspace{\SubBeforeNeg}
\subsection{Layer-wise Representation Probes}
\vspace{\SubAfterNeg}
\label{sec:repr_div}

\vspace{\ParaVspace}
For each condition, we save the projector output, decoder residual streams, and audio-token span $[s,e)$ from the generation-regime forward. At layer $\ell$, we mean-pool the span, $\bar{\mathbf{h}}^{({\ell,c})}=\frac{1}{e-s}\sum_{t=s}^{e-1}\mathbf{h}^{({\ell,c})}_{[t,:]}$, and define per-band divergence $\Delta_{\text{llm}}^{({\ell})}(b_k)=\frac{1}{n}\sum_i\|\bar{\mathbf{h}}_i^{({\ell,\text{adv}})}-\bar{\mathbf{h}}_i^{({\ell,b_k})}\|_2$. We correlate band-level ASR drops with these divergences at each layer, $\rho_\ell=\mathrm{corr}_k\big(\mathrm{ASR\_Drop}(b_k),\Delta^{(\ell)}_{\mathrm{llm}}(b_k)\big)$ (Pearson; Spearman is a sensitivity analysis). This raw correlation is descriptive and uses only eight bands. We summarize it with the late-decoder ($\ell\ge15$) slope. A separate prompt-by-band model estimates the association after adjustment for prompt identity and band energy (Section~\ref{sec:meth_inference}).

\vspace{\SubBeforeNeg}
\subsection{Statistical Inference and Uncertainty}
\vspace{\SubAfterNeg}
\label{sec:meth_inference}

\vspace{\ParaVspace}
We validate the primary judge on 220 condition-blind, human-labeled responses: 120 adversarial responses, 40 clean controls, and 60 responses sampled from disagreement strata. One author annotated the set; a second annotator was unavailable (Section~\ref{sec:limitations}). Appendix~\ref{sec:app_judge} gives the sample composition, seed, and blind key. On the 120 adversarial responses, the judge's false-positive rate is $0.095$ ($2/21$ human negatives; Wilson 95\% CI $[0.027,0.289]$), and its false-negative rate is $0.131$ ($13/99$ human positives; Wilson 95\% CI $[0.078,0.212]$). Applying the Rogan-Gladen correction to the $n=520$ judged adversarial ASR of $0.767$ gives approximately $0.87$; propagating the endpoint confidence intervals gives approximately $0.83$--$0.95$. These error rates were estimated only for adversarial responses, so we do not apply the correction to masked conditions. For example, applying the adversarial false-positive rate to the judged $b_8$ ASR of $0.05$ gives a negative corrected rate. We therefore report the uncorrected judge scores for condition-level comparisons and treat the corrected adversarial estimate as a sensitivity analysis. The primary judge uses the AdvWave LLM-as-a-judge prompt, its default threshold, and GPT-4o-mini at temperature $0$ (Section~\ref{sec:exp_setup}).

\vspace{\ParaBeforeNeg}
\paragraph{Raw depth curve.} The raw statistic $\rho_\ell$ is a correlation across the $K$ bands at layer $\ell$; we summarize its late-decoder slope over $\ell\in[15,31]$. Adjacent layers are strongly autocorrelated (effective $n\approx4.3$ over 17 layers). We compare the observed curve with phase-randomized surrogates that preserve its power spectrum~\citep{theiler1992surrogate,schreiber2000surrogate}; the observed slope does not reject this null ($p=0.183$). Separately, a prompt-level bootstrap over 2,000 resamples at $n=520$ estimates uncertainty in $\Delta\rho=\rho_{30}-\rho_0$. This interval is conditional on the eight-band partition and does not replace the surrogate test.

\vspace{\ParaBeforeNeg}
\paragraph{Energy-adjusted depth model and partition sensitivity.} Because band energy and representational divergence are strongly correlated, we estimate their relationship at the prompt-by-band level rather than relying on eight band means. We restrict the analysis to the 399 prompts jailbroken by the full adversarial input and, at each layer $\ell$, fit a mixed-effects logistic model: $\mathrm{logit}\,P(\mathrm{mask\ breaks\ attack}{=}1)=\alpha+\beta_\ell\log E_{p,k}+\gamma_\ell\Delta_{\mathrm{llm},p,k}^{(\ell)}+u_p$, with a random intercept for each prompt. Energy is measured per prompt and band, and divergences are normalized per layer. We report $\gamma_\ell$ with a prompt-cluster bootstrap CI, focusing on the pre-specified layer-30 coefficient and the layer-30-minus-projector contrast (Section~\ref{sec:res_repr}). The raw band-level result remains sensitive to the partition: its leave-one-band-out spread is $0.172$, about twice the prompt-bootstrap standard error. This sensitivity applies to the raw eight-band correlation, not the prompt-level model's uncertainty interval.

\vspace{\SecBeforeNeg}
\section{Experimental Setup}
\label{sec:experiments}
\label{sec:exp_setup}

\paragraph{Victim models, attack, and corpus.}
We evaluate Qwen2-Audio-7B-Instruct~\citep{chu2024qwen2audio} (Whisper-large-v3 encoder, $L{=}32$, $d_{\text{llm}}{=}4096$) and LLaMA-Omni~\citep{fang2025llamaomni}. Both models use a $128$-bin log-mel frontend, distinct from the Qwen analysis basis described in Section~\ref{sec:setup}. We attack 520 prompts from AdvBench~\citep{zou2023universal} with AdvWave-P~\citep{jin2025almguard}, the additive variant of AdvWave~\citep{kang2025advwave}. The perturbation is applied to the full waveform; its budget and optimization settings appear in Section~\ref{sec:setup} and Figure~\ref{fig:methodology}. The primary $K{=}8$ mask and keep-only sweeps and the depth estimators use all 520 prompts. The $K{=}16$ behavioral sweep also uses 520 prompts, but its hidden-state depth estimate uses $n{=}50$. Several secondary controls, including equal-Hz and equal-energy partitions, scattered removal, and global rescaling, are reported at their respective sample sizes and are described separately from the primary-judge results. The epsilon and resolution-ladder analyses use $n{=}50$. Main sweeps use greedy decoding and one response per prompt and condition. Stochastic-decoding results are a sensitivity check, not headline ASR (Appendix~\ref{sec:app_seed_stability}). The mask and keep-only sweeps are separate, each with $K{+}2$ conditions: clean, adversarial, and $K$ masked or retained variants. Representation visualizations, the Omni arm, and the transfer matrix use the conditions specified in their captions. Omni's clean ASR leaves limited headroom for measuring per-band effects. For transfer, each source perturbation is passed unchanged to the target, and source bands are swept in each cell of the $2\times2$ ASR matrix.

\vspace{\ParaBeforeNeg}
\paragraph{Judge.}
For the primary $K{=}8$ results, we use the AdvWave LLM-as-a-judge prompt~\citep{kang2025advwave}, its default threshold, and GPT-4o-mini at temperature $0$. The provider did not expose an immutable model snapshot; the main $n=520$ scores were collected through 2026-09-02, and per-sample responses and scores are needed to reproduce these labels if the model changes. Equal-Hz and equal-energy partitions, scattered removal, graded attenuation, global rescaling, and the ALMGuard suppression sweep were scored with Meta-Llama-3-8B-Instruct. The GRM re-optimization pilot uses GPT-4o-mini. Each result is interpreted within its scorer-specific run, and ASRs from different judges are not pooled. We also scored the $n=50$ $K{=}8$ sensitivity set with the proprietary reference model GPT-5.6-sol (scored 2026-09-05); this provider model label has no archived immutable snapshot. The reference scores appear in Table~\ref{tab:judge_agreement} and are a sensitivity analysis, not the headline ASR. The clean-audio renderer is described in Appendix~\ref{sec:appendix_renderer}.

\vspace{\SecBeforeNeg}
\section{Results and Discussion}
\vspace{\SecAfterNeg}
\label{sec:results}

\begin{table}[t]
\centering
\small
\setlength{\tabcolsep}{4pt}
\renewcommand{\arraystretch}{1.05}
\newcolumntype{Q}{>{\raggedright\arraybackslash}p{3.7cm}}
\newcolumntype{T}{>{\raggedright\arraybackslash\hsize=0.85\hsize}X}
\newcolumntype{A}{>{\raggedright\arraybackslash\hsize=1.15\hsize}X}
\begin{tabularx}{\linewidth}{@{} Q T A @{}}
\toprule
\textbf{Question} & \textbf{Test} & \textbf{Answer} \\
\midrule
Is one frequency band privileged? (\S\ref*{sec:res_asr}) & $K{=}8$ mask sweep; equal-Hz/energy partitions; $K{=}16$ equal-energy result & \textbf{Unresolved across resolutions.} Energy share predicts the $K{=}8$ ranking ($\rho_s{=}0.95$); the $K{=}16$ top band is unresolved without a matched control \\
Does removal structure matter? (\S\ref*{sec:res_fidelity}) & Contiguous mask vs.\ matched-energy scattered removal; global-rescaling check & \textbf{Yes, in the tested comparisons.} Contiguous removal is more damaging in lower bands; both methods approach the floor in upper bands. Rescaling tests amplitude, not location \\
Do restricted supports sustain the attack? (\S\ref*{sec:res_asr}) & Keep-only; cumulative addition; band-restricted re-optimization & \textbf{Some random supports do.} Tested single bands and their union perform poorly; random supports covering about half the STFT bins reach mean ASR $0.86$ in an $n{=}20$ pilot \\
Do prior band selections outperform controls? (\S\ref*{sec:res_fidelity}) & ALMGuard mask and GRM rule vs.\ matched random & \textbf{Inconclusive.} Both pilots are limited in scale and do not test the methods' original objectives \\
Do band importance and representation covary? (\S\ref*{sec:res_repr}) & Raw $\rho_\ell$ curve; energy-adjusted prompt-by-band model & \textbf{Yes, with different evidence.} The raw increase is partition-sensitive and does not reject its surrogate null; the adjusted association is positive across depth \\
Does the adjusted association change with depth? (\S\ref*{sec:res_repr}) & Prompt- and energy-adjusted mixed-effects model; behavior-irrelevant controls & \textbf{The estimate increases.} The layer-30-minus-projector contrast is $+0.293$ (95\% CI $[0.11,0.51]$); this is not causal localization \\
What replicates across corpora? (\S\ref*{sec:res_supplemental}) & Full JailbreakBench pool; selected MaliciousInstruct and StrongREJECT subsets & \textbf{Limited evidence.} The full JailbreakBench sweep reproduces the broad ranking; the other subsets are selection-conditioned. None rejects the depth surrogate null \\
\bottomrule
\end{tabularx}
\vspace{-0.4cm}
\caption{Summary of findings (Qwen2-Audio, AdvWave-P, $\epsilon{=}0.01$, $n{=}520$ unless noted).}\label{tab:summary}
\end{table}

\vspace{\ParaVspace}
On Qwen2-Audio, the clean-input ASR is $0.008$ and the adversarial ASR is $0.767$ across the 520 AdvBench prompts. The condition-blind judge validation implies a corrected adversarial estimate of about $0.87$, with a wide interval of approximately $0.83$--$0.95$; we do not transport this correction to masked conditions. Unless noted otherwise, the primary behavioral results use GPT-4o-mini, and the $95\%$ intervals are prompt-bootstrap intervals. Secondary control sweeps and their scorers are identified in Section~\ref{sec:exp_setup}. Table~\ref{tab:summary} collects the main findings.

\vspace{\SubBeforeNeg}
\subsection{Per-Band Behavioral Importance}
\vspace{\SubAfterNeg}
\label{sec:res_asr}

\begin{figure}[t]
  \centering
  \includegraphics[width=0.9\textwidth, trim=40 20 40 15]{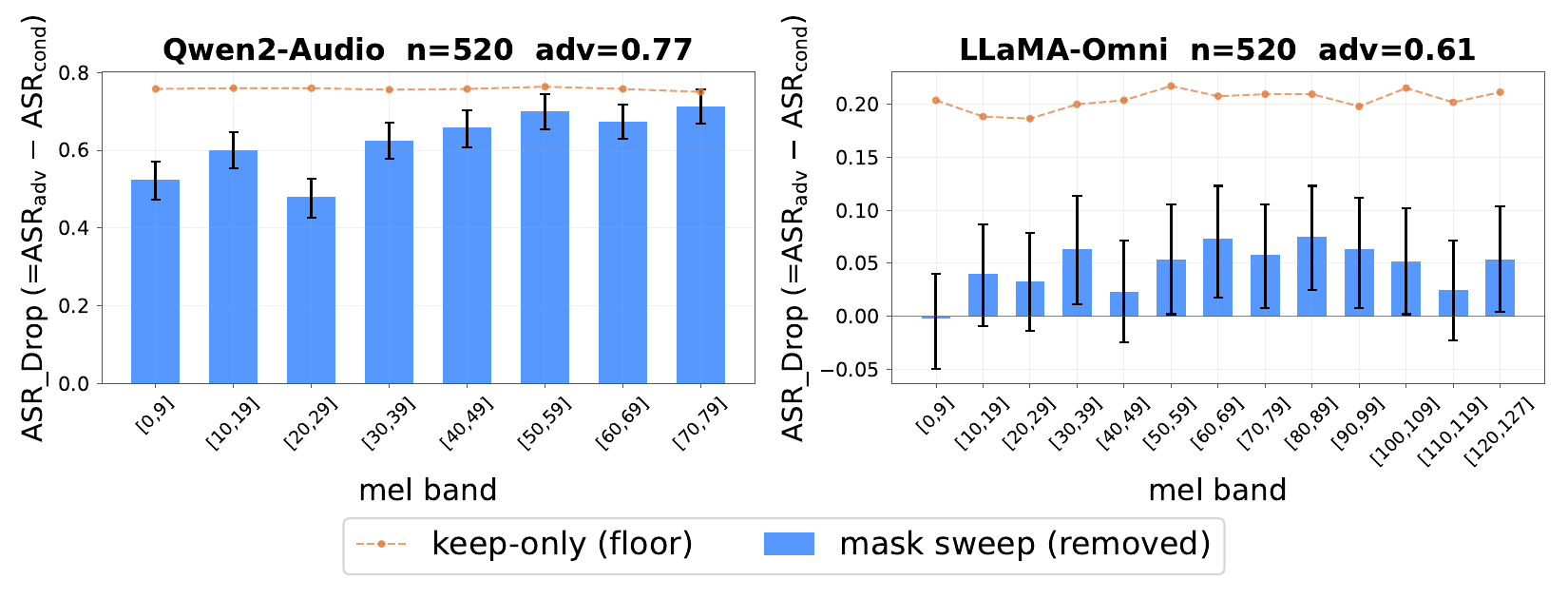}
  \vspace{-0.2cm}
  \caption{Per-band behavioral importance on the $n{=}520$ corpus; bars show mask-sweep and keep-only drops, with prompt-bootstrap 95\% CIs for the primary mask sweep. On Qwen2-Audio (left, judged adversarial ASR $0.77$), upper-band removal sharply reduces ASR in the $K{=}8$ partition. This ranking is not stable across partitions: energy share predicts it closely, equal-energy $K{=}8$ masks all leave low residual ASR, and the $K{=}16$ top-band result remains unresolved without a matched control. The keep-only sweep leaves ASR near the floor when retaining one band's original perturbation components; re-optimized support tests are reported separately. On LLaMA-Omni (right, judged adversarial ASR $0.61$), the mask-drop CIs overlap zero.}
  \label{fig:qwen2_asr_mask_vs_keep}
\end{figure}

\vspace{\ParaVspace}
Figure~\ref{fig:qwen2_asr_mask_vs_keep} reports the primary $K{=}8$ sweeps at $n{=}520$. Masking the top band $b_8{=}[70,79]$ reduces judged ASR from $0.767$ to $0.054$ (drop $0.714$, 95\% CI $[0.671,0.754]$). The four upper bands $b_5$--$b_8$ span $1.7$--$8$\,kHz, or about $79\%$ of the analyzed spectrum. A separate $n{=}50$ control removed a random contiguous region covering 40 of the 80 analysis bins, the same bin count as $b_5$--$b_8$ combined. Across 20 draws, mean ASR was $0.015$, compared with $0.05$ after masking $b_8$. Both results are near the floor, so this broad-region control does not distinguish a specific location. The equal-energy partitions and energy-share analysis below provide more informative comparisons. The keep-only sweep also leaves ASR near the floor when retaining a single band's original perturbation components; the re-optimized support pilot is reported separately.

\vspace{\ParaBeforeNeg}
\paragraph{Resolution controls.} In the equal-Hz $K{=}8$ sweep, masking any band leaves residual ASR at or below $0.30$. The perturbation's energy is unevenly distributed: frequencies below $1$\,kHz contain $23.0\%$ of its energy. We therefore also partition the spectrum into eight contiguous bands with equal shares of corpus-mean $|\mathrm{STFT}(\delta)|^2$ energy. In this secondary sweep, masking each band leaves residual ASR between $0.07$ and $0.16$; the range does not justify calling the bands equally important. At the finer, 16-band equal-energy resolution, masking the narrow top band $b_{15}{=}[7520,7960]$\,Hz leaves residual ASR $0.10$ (drop $0.667$; energy share $0.065$). This is the largest observed drop at that resolution, but there is no matched random control, so its frequency-specific interpretation remains unresolved (Section~\ref{sec:limitations}). The results therefore show both partition sensitivity at $K{=}8$ and an unresolved top-band result at $K{=}16$, rather than a resolution-independent frequency ranking. In the separate $n{=}50$ cumulative-retention analysis, the importance- and energy-ordered curves are similar: ASR reaches $0.28$ at $k{=}7$ and $0.80$ at $k{=}8$, the full adversarial level for that subset (Figure~\ref{fig:r2_cumulative}).

\vspace{\SubBeforeNeg}
\subsection{Protocol Diagnostics: Occlusion Fidelity and the Energy Confound}
\vspace{\SubAfterNeg}
\label{sec:res_fidelity}

\vspace{\ParaVspace}
Occlusion fidelity and band-energy diagnostics appear in Appendix~\ref{sec:app_energy_diag}. Reconstruction residuals and neighboring-band leakage are small relative to the removed energy. In the $n{=}520$ matched-energy scattered control, each random comparison removes the same number of STFT bins and the same total perturbation energy as its contiguous-band counterpart. Across 20 draws per band, contiguous removal leaves lower ASR in every band; the difference is largest in lower bands and narrows toward the top, where both conditions approach the floor. This indicates that removal structure matters, but does not by itself identify one uniquely necessary band. Graded attenuation at 3, 6, and 12 dB produces a monotone response. Raw audio-span divergence also tracks band energy closely (Appendix~\ref{sec:app_energy_diag}), motivating the prompt- and energy-adjusted model in Section~\ref{sec:res_repr}.

\vspace{\ParaBeforeNeg}
\paragraph{Global rescaling checks amplitude sensitivity.} For each $K{=}8$ band, a secondary control scales the full perturbation by $\sqrt{1-e_k}$, where $e_k$ is that band's energy share, while preserving its direction. Under the Meta-Llama-3-8B-Instruct judge, ASR remains near the adversarial baseline ($0.75$--$0.79$; uniform $\tfrac18$ energy removal gives $0.76$). This is an amplitude-sensitivity check, not a location control: the mask displacement is $\sqrt{e_k}=0.27$--$0.49$ in relative $L_2$ distance, while rescaling changes the perturbation by only $1-\sqrt{1-e_k}=0.037$--$0.13$. The global-rescaling and mask edits therefore differ in displacement as well as frequency selectivity. The separate scattered-removal comparison matches the number of bins and removed energy to test removal structure.

\vspace{\ParaBeforeNeg}
\paragraph{Named prior band selections vs.\ matched random.} We tested the selected supports from ALMGuard and GRM on AdvWave-P, not the methods' original objectives (Appendix~\ref{sec:app_audit}). With equal-energy suppression at $n{=}520$, ALMGuard's mask gives ASR $\AlmAsr$; the 20 matched random masks have mean $\AlmRandMean$ and range $[0.150,0.187]$. This pilot is inconclusive under the pre-specified comparison. In a separate $n{=}20$ re-optimization pilot, GRM's selected support gives ASR $\GrmAsr$, compared with a mean of $\GrmRandMean$ for bandwidth-matched random supports (range $[\GrmRandLo,\GrmRandHi]$). GRM lies at the lower edge of this range, so this small pilot is also inconclusive. Neither test establishes whether the source methods' selections are effective for their original objectives.

\vspace{\SubBeforeNeg}
\subsection{Per-Layer Representational Importance and the Alignment Trend}
\vspace{\SubAfterNeg}
\label{sec:res_repr}

\vspace{\ParaVspace}
The raw $K{=}8$ correlation rises from $\rho_0=+0.568$ at the projector output to $\rho_{30}=+0.877$ at layer 30 (Figure~\ref{fig:r12_full_depth}). The paired prompt bootstrap estimates $\Delta\rho=+0.307$ (95\% CI $[0.271,0.342]$), conditional on this partition. The raw curve does not reject the phase-randomized surrogate null ($p=0.183$), and its leave-one-band-out spread is $0.172$, about twice the prompt-bootstrap standard error. These results support an increase over prompts for the fixed partition, but do not establish that the raw trend is specific to jailbreak behavior. The pooled curve rises with depth, while the final-audio-frame correlation stays near $0.45$--$0.49$. A length- and position-matched random-token span reproduces the pooled level, further limiting the interpretation of the raw curve. Table~\ref{tab:rho_estimators} summarizes the raw estimators and the adjusted model.

\vspace{\ParaBeforeNeg}
\paragraph{Interpreting the raw depth curve.} The prompt-bootstrap interval excludes zero for the raw increase, but this inference is conditional on eight bands. Its leave-one-band-out spread is $0.172$, and the phase-randomized test does not reject an autocorrelation-preserving null ($p=0.183$). The prompt- and energy-adjusted model addresses a different question and is reported separately. Single-layer activation patching restores the jailbreak only near the decoder entrance (Section~\ref{sec:res_causal}). Because later positions have already processed the occluded audio at lower layers, this design cannot test representations that develop cumulatively; it does not establish a causal layer.

\vspace{\ParaVspace}
\textbf{The adjusted association increases with depth.} Among the 399 prompts jailbroken by the full adversarial input, the prompt-by-band mixed-effects model estimates a positive divergence coefficient at every layer, rising from $\AGammaZero$ at the projector output to $\AGammaThirty$ at layer 30 (95\% CI $[\AGammaThirtyLo,\AGammaThirtyHi]$). The pre-specified layer-30-minus-projector contrast is $\ADepthDelta$ (95\% CI $[0.11,0.51]$), and the mean late-layer coefficient is $\ASlopeLate$ (95\% CI $[\ASlopeLateLo,\ASlopeLateHi]$). The tested behavior-irrelevant band statistics have zero depth contrasts. This supports a depth-varying, energy-adjusted association in this attack and partition, not a causal layer location. The controls share the same coarse frequency partition, and single-layer activation patching cannot test effects that develop cumulatively across layers. The raw correlation and adjusted model answer different questions: the former is partition-sensitive and does not reject its surrogate null, while the latter controls for prompt identity and per-prompt band energy.

\vspace{\SubBeforeNeg}
\subsection{Activation Patching Is a Null for Localization}
\vspace{\SubAfterNeg}
\label{sec:res_causal}
\vspace{\ParaVspace}
We test single-layer activation patching in raw-overwrite and norm-matched modes~\citep{zhang2024bestpractices}. The self-patch gate passes all 312 checks. Restoration decreases with depth, from $0.83$ at the projector output to zero at layers $18$--$30$; the control band shows a similar pattern. Patching the full adversarial span into the clean stream restores the attack only near the decoder entrance. This intervention replaces the full audio span and cannot isolate a specific band. With $n{=}12$, the result is a design-limited null, not evidence for where the representation forms (Appendix~\ref{sec:app_activation}).

\vspace{\SubBeforeNeg}
\subsection{Additional Findings: Geometry, Transfer, Detector, and Robustness}
\vspace{\SubAfterNeg}
\label{sec:res_supplemental}

\vspace{-0.1cm}
\paragraph{Geometry, robustness, and detection.} A supervised input-space classifier matches the depth probe's detection AUROC, so these experiments show no measurable detection advantage for residual-stream features (Appendix~\ref{sec:app_detector}). Gradient attribution ($\rho=0.97$, $p<0.001$) and perturbation-energy share ($\rho=0.95$) both recover the standard per-band ranking. These baselines limit the ranking's novelty; the value of the occlusion analysis is in testing it against partition, energy, and removal-structure controls. Masking the low band $b_1{=}[0,9]$ reduces ASR from $0.77$ to $0.24$. Appendix~\ref{sec:app_seed_stability} reports decoding- and optimization-seed checks. Appendix~\ref{sec:app_defense} explains why this perturbation-only notch is not a validated defense and may suppress useful speech.

\vspace{\ParaBeforeNeg}
\paragraph{Cross-corpus analyses.} Only the JailbreakBench $K{=}8$ mask sweep uses the full jailbroken pool without selection on band necessity ($n{=}70$ of 89). The MaliciousInstruct and StrongREJECT mask analyses use the existing 50-prompt, highest-necessity subsets, selected from jailbroken pools of 93/100 and 190/276, respectively (Table~\ref{tab:second_corpus_mask}; Appendix~\ref{sec:app_second_corpus}). The JailbreakBench full-pool sweep reproduces the broad pattern that masking $70$--$79$ has the largest drop and masking $20$--$29$ the smallest. The other two subsets are selection-conditioned and do not independently replicate that ranking. Depth estimates use existing $n{=}50$ highest-necessity samples for all three corpora, including JailbreakBench; none rejects the phase-randomized null ($p{=}\PJbbP$, $\PMiP$, $\PSrP$; primary $p{=}\DepthSurrP$). The separate $n{=}20$ re-optimization pilot shows that some random supports covering about half the STFT bins sustain the attack; it does not establish a general sufficiency threshold.

\vspace{\ParaBeforeNeg}
\paragraph{LLaMA-Omni and cross-model transfer.} On the $n{=}520$ corpus, Omni's judged clean ASR is $0.41$ and adversarial ASR is $0.61$, leaving limited headroom for per-band comparisons (Appendix~\ref{sec:app_omni}). Direct cross-model transfer is a two-directional null; these tests identify no transferable band or depth signature (Appendix~\ref{sec:app_transfer}).

\vspace{\SecBeforeNeg}
\section{Conclusion}
\vspace{\SecAfterNeg}
\label{sec:conclusion}

\vspace{\ParaVspace}
We audited AdvWave-P on Qwen2-Audio using frequency occlusion and decoder representations. The standard eight-band ranking is closely predicted by band energy, and its apparent location changes with the analysis partition. A finer equal-energy sweep singles out a narrow band near Nyquist, but the absence of a matched control leaves that result unresolved. Matched-energy scattered-removal tests show that contiguous removal is more damaging in lower bands; at higher bands, both conditions approach the floor. Global rescaling leaves ASR near baseline, but it is an amplitude check rather than a frequency-location control. In re-optimization pilots, tested single-band supports and their union perform poorly, while random supports covering about half the STFT bins reach mean ASR $0.86$. The ALMGuard and GRM audits are inconclusive and do not evaluate those methods' original objectives. A prompt- and energy-adjusted model estimates a depth-varying association between audio-span divergence and band necessity, but the raw depth trend is partition-sensitive and single-layer patching does not establish a causal location. Overall, the results motivate partition-aware controls and leave open whether the findings generalize to other attacks, victims, or perturbation domains.

\vspace{\SecBeforeNeg}
\section{Limitations}
\vspace{\SecAfterNeg}
\label{sec:limitations}

\paragraph{Scope.} The positive findings come from one attack, AdvWave-P at $\epsilon=0.01$, on one victim with a usable margin, Qwen2-Audio. LLaMA-Omni already judges $40\%$ of clean harmful prompts as jailbroken, leaving too little headroom to measure per-band effects. The protocol is designed for additive perturbations aligned with the clean waveform; it does not directly apply to suffix-appended, semantic, or mel-domain attacks.

\vspace{\ParaBeforeNeg}
\paragraph{Perturbation budget.} Attacks optimized at $\epsilon=0.005$ and $0.02$ were not stable under cold-start re-inference. Their ASRs fell from $0.84$ and $0.76$ during optimization to $0.32$ and $0.40$ at inference on $n=50$ prompts; we have not isolated the cause. All behavioral results therefore use $\epsilon=0.01$. The untrained-$\epsilon$ probe detects the perturbation family, not necessarily a successful jailbreak.

\vspace{\ParaBeforeNeg}
\paragraph{Occlusion target.} We remove components from the perturbation while holding the clean waveform fixed. This isolates perturbation structure but does not test a deployable filter, which would have to process the clean-plus-adversarial mixture. The notch result in Appendix~\ref{sec:app_defense} should be read as a perturbation analysis, not as a validated defense.

\vspace{\ParaBeforeNeg}
\paragraph{Removal displacement.} Global rescaling preserves the perturbation's direction but does not match the band mask's displacement. For a band with energy share $e_k$, the mask's relative $L_2$ displacement is $\sqrt{e_k}$ ($0.27$--$0.49$), whereas the rescaling displacement is $1-\sqrt{1-e_k}$ ($0.037$--$0.13$), 3.8--7.3 times smaller (Table~\ref{tab:protocol_energy}). We therefore treat rescaling only as an amplitude check. The comparison of removal structure relies on the scattered control, which matches both removed energy and bin count.

\vspace{\ParaBeforeNeg}
\paragraph{Equal-energy resolution.} In the $K=8$ equal-energy partition, residual ASR ranges from $0.07$ to $0.16$ at $n=520$. This spread is several per-band standard errors, so the result supports the claim that masking any band disrupts the attack, not that all bands are equally important. At $K=16$, the narrow near-Nyquist band $b_{15}=[7520,7960]$ Hz leaves residual ASR $0.10$ and carries the largest energy share. Without a matched random control at this resolution, we treat this as an unresolved observation, not a localization result.

\vspace{\ParaBeforeNeg}
\paragraph{Depth.} The adjusted depth result is an association, not a causal localization. The raw correlation uses only eight bands; its prompt-bootstrap interval excludes zero, but the leave-one-band-out spread is $0.172$, about twice the bootstrap standard error, and the phase-randomized null test gives $p=0.183$. These raw-curve results are conditional on the partition and do not test the adjusted model. The prompt-by-band model adjusts for prompt identity and per-prompt energy; its layer-30-minus-projector contrast is positive, while the tested behavior-irrelevant statistics show no corresponding depth contrast. Those controls use statistics defined on the same coarse partition. Single-layer activation patching cannot test effects that develop cumulatively across layers, so the causal location remains unresolved. Cumulative or KV-cache patching, a refusal-direction readout~\citep{arditi2024refusal}, and a check for massive-activation channels~\citep{sun2024massive} are possible next steps.

\vspace{\ParaBeforeNeg}
\paragraph{Judge.} We validated GPT-4o-mini's AdvWave-judge labels against one author-annotator on a condition-blind sample; a second annotator was unavailable. Applying the Rogan-Gladen correction to the adversarial condition gives $0.87$, with an approximate interval of $0.83$--$0.95$ after propagating uncertainty in the validation rates. It cannot be transferred to masked conditions: applying the adversarial false-positive rate of $0.095$ to the judged $b_8$ ASR of $0.05$, for example, yields a negative corrected rate. We therefore report uncorrected scores for condition-level comparisons. The Wilson 95\% CIs are $[0.03,0.29]$ for the false-positive rate ($2/21$) and $[0.08,0.21]$ for the false-negative rate ($13/99$). The $n{=}50$ reference-judge sensitivity check uses GPT-5.6-sol, a provider model label for which no immutable snapshot was archived; it gives adversarial ASR $0.64$ versus $0.76$ under the primary judge. The ranking agreement on this sensitivity subset does not remove uncertainty about absolute ASR or about judge-version drift.

\vspace{\ParaBeforeNeg}
\paragraph{Cross-corpus analyses.} These studies use a different TTS renderer. Only the JailbreakBench $K{=}8$ mask sweep covers its full jailbroken pool ($n=70$); MaliciousInstruct and StrongREJECT use the 50 highest-necessity prompts from their respective jailbroken pools. The selected subsets cannot independently validate the ranking used to select them. Depth curves for all three corpora use the existing 50-prompt highest-necessity samples, not the full JailbreakBench pool, and none rejects the phase-randomized null. Re-optimized sufficiency and the GRM support comparison are small pilots ($n=20$); in the GRM audit, random supports are scattered and are not structurally matched to the selected support (Appendix~\ref{sec:app_audit}).

\vspace{\SecBeforeNeg}
\section*{Ethical Considerations}
\vspace{\SecAfterNeg}
\label{sec:ethics}

\vspace{\ParaVspace}
\textbf{Use and release of resources.} We apply the public AdvWave-P method to public harmful-prompt corpora and public victim models; we introduce no new attack method, prompt corpus, or model. The optimized adversarial waveforms and perturbations are not included in the release. Instead, the repository will provide optimization and analysis code, seeds, hyperparameters, renderer details, prompt identifiers, responses, and judge scores to support re-derivation without distributing those waveforms (Appendix~\ref{sec:appendix_artifacts}). The analysis is intended to inform defense, though frequency-level findings may also help characterize existing attacks. We recruited no external participants and trained no new models. The experiments used fewer than $100$ A100 GPU-hours.

\vspace{\SecBeforeNeg}
\section*{Acknowledgment}
\vspace{\SecAfterNeg}
\vspace{-0.1cm}
This work was supported in parts by the NYUAD Center for Cyber Security (CCS), funded by Tamkeen under the NYUAD Research Institute Award G1104. Experiments are performed with NYUAD Jubail High Performance Computing (HPC).

\bibliographystyle{template/iclr2027_conference}
\bibliography{reference}

\clearpage

\appendix

\vspace{\SecBeforeNeg}
\section{Appendix}
\vspace{\SecAfterNeg}
\label{sec:appendix}

\subsection{Methodology and Protocol}
\vspace{\SubAfterNeg}
\label{sec:app_meth}

\vspace{\ParaVspace}
Figure~\ref{fig:methodology} summarizes the analysis. We decompose the perturbation $\bm\delta$ in the STFT domain, remove or retain one band, reconstruct the waveform, and evaluate it with the victim and judge. We then compare each band's behavioral importance with audio-span divergence across decoder layers.

\vspace{\SubBeforeNeg}
\subsection{Representation-Space Visualizations}
\vspace{\SubAfterNeg}
\label{sec:appendix_repr}

\vspace{\ParaVspace}
We supplement Section~\ref{sec:res_supplemental} with linear CKA heatmaps for the $K{+}2$ clean, adversarial, and band-masked conditions at layer 30, plus PCA and t-SNE projections of the pooled audio-span states.

\begin{figure}[b]
  \centering
  \includegraphics[width=0.48\linewidth]{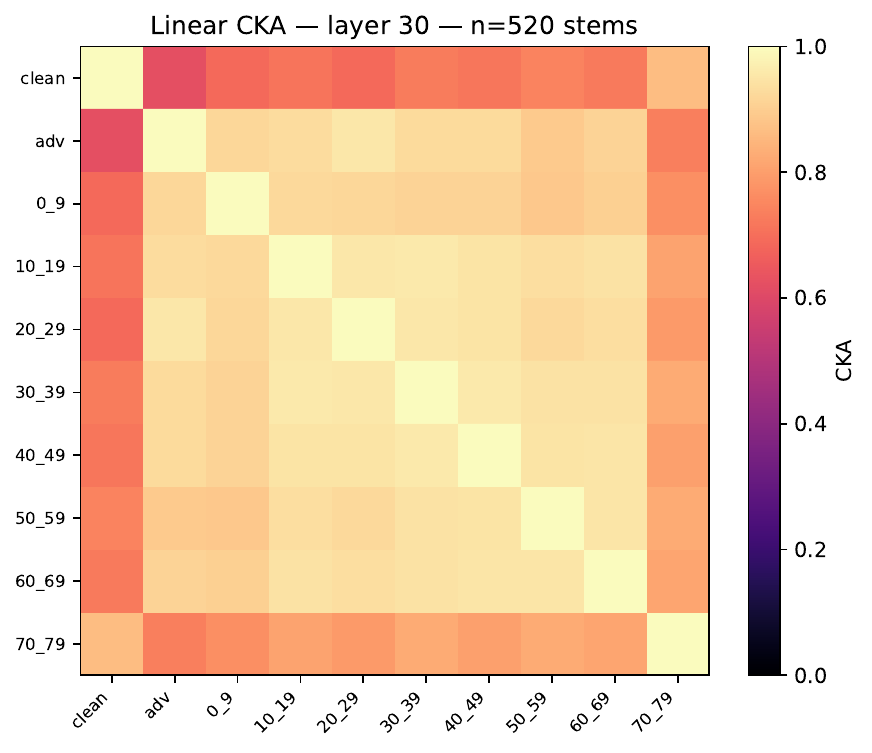}\hfill
  \includegraphics[width=0.48\linewidth]{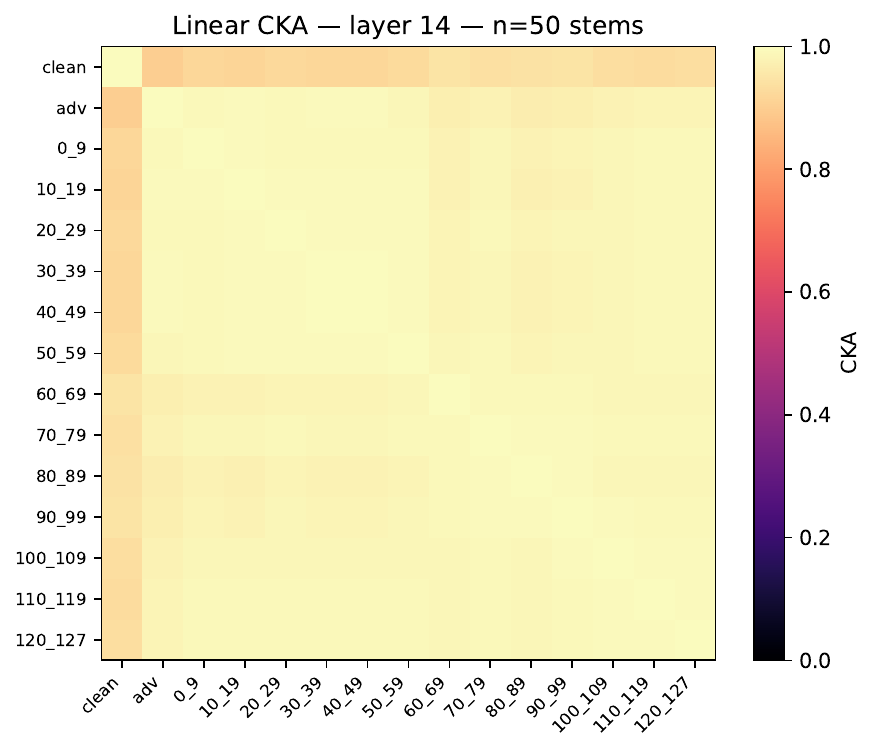}
  \vspace{-0.4cm}
  \caption{Linear CKA (Qwen2-Audio at $\ell{=}30$, $n{=}520$; LLaMA-Omni at $\ell{=}14$, $n{=}50$ pairing, which lacks the $n{=}520$ sampling point at $\ell{=}14$). \textbf{Left:} Qwen2-Audio ($10{\times}10$). \textbf{Right:} LLaMA-Omni ($15{\times}15$). Cells are linear CKA between audio-span-pooled residual states \citep{kornblith2019linear}. Band-removed conditions form a broad high-similarity block, so the behaviorally relevant differences in Figure~\ref{fig:qwen2_rho} live in the high-dimensional distance rather than in CKA-detectable geometry.}
  \label{fig:cka_qwen2_l30}
\end{figure}

\begin{figure}[tbp]
  \centering
  \includegraphics[width=0.48\linewidth]{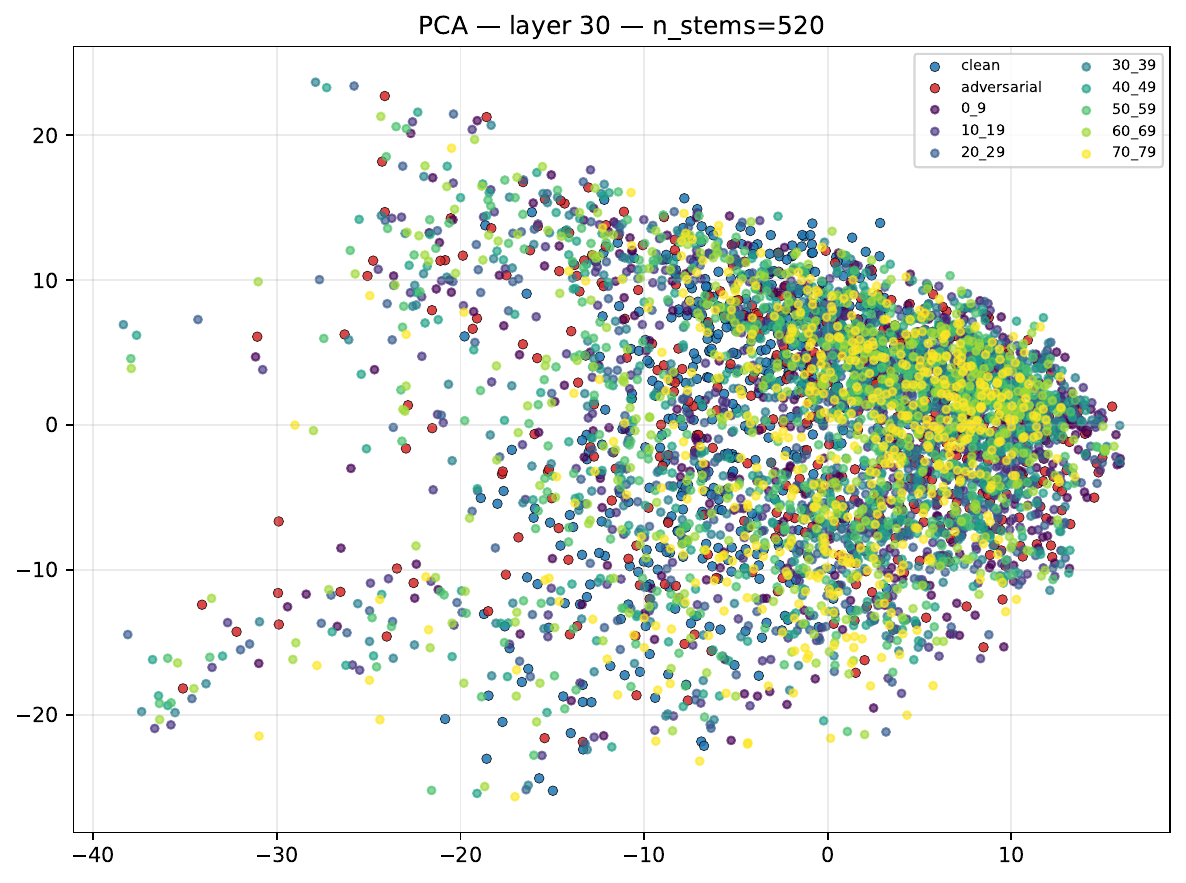}\hfill
  \includegraphics[width=0.48\linewidth]{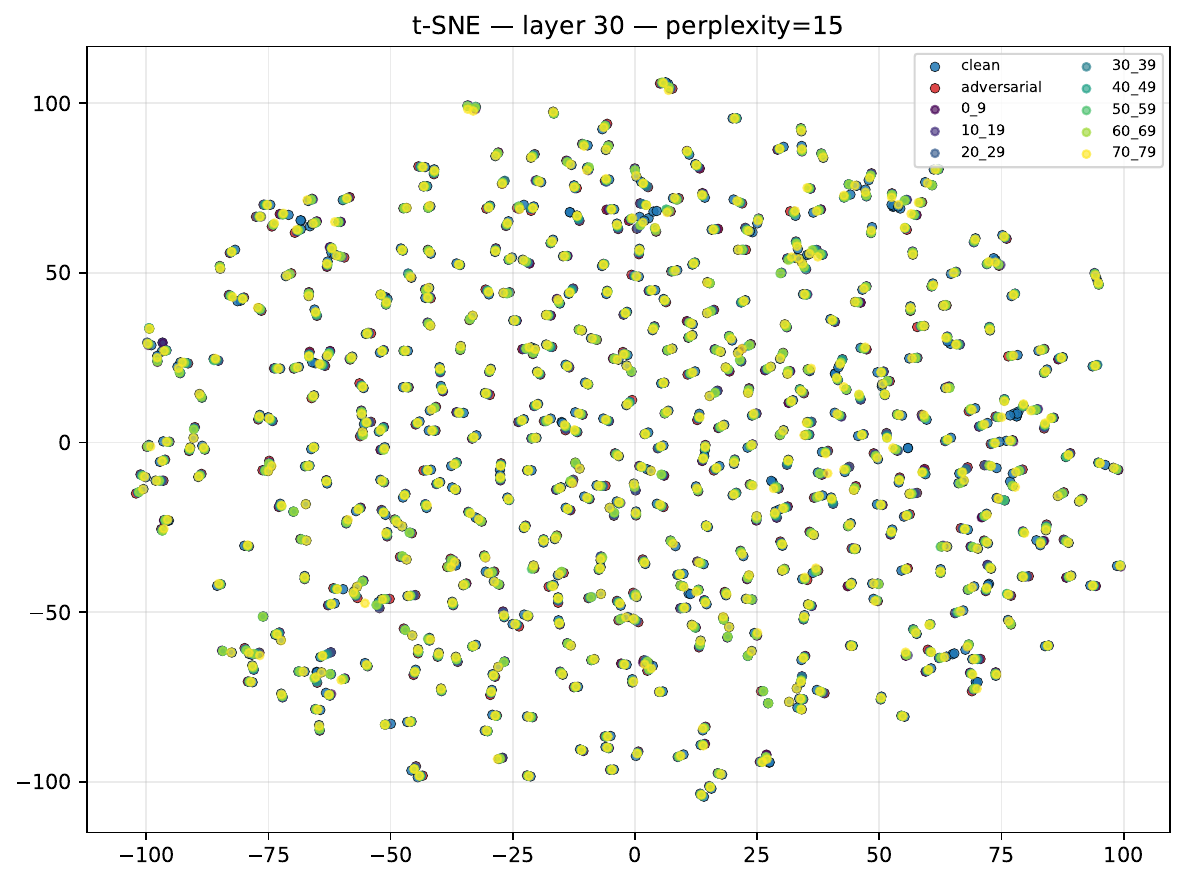}
  \vspace{-0.3cm}
  \caption{Qwen2-Audio at layer $30$, $n{=}520$. \textbf{Left:} PCA. \textbf{Right:} t-SNE (perplexity $15$). Each point is one (sample, condition) pair across clean, full adversarial, and $K{=}8$ band-removed variants. Conditions overlap in both projections, confirming the jailbreak direction is not on the first two unsupervised axes.}
  \label{fig:pca_tsne_qwen2_l30}
\end{figure}

\begin{figure}[b]
  \centering
  \includegraphics[width=0.48\linewidth]{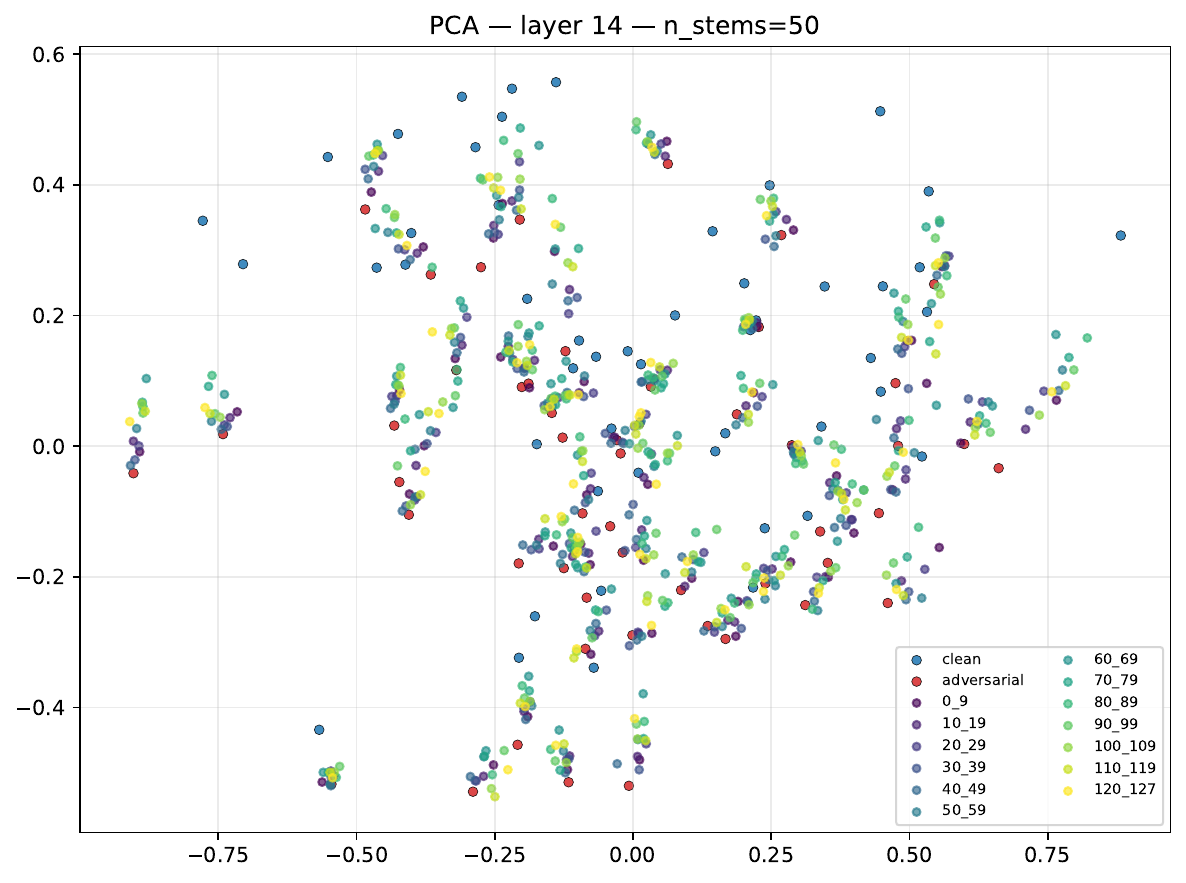}\hfill
  \includegraphics[width=0.48\linewidth]{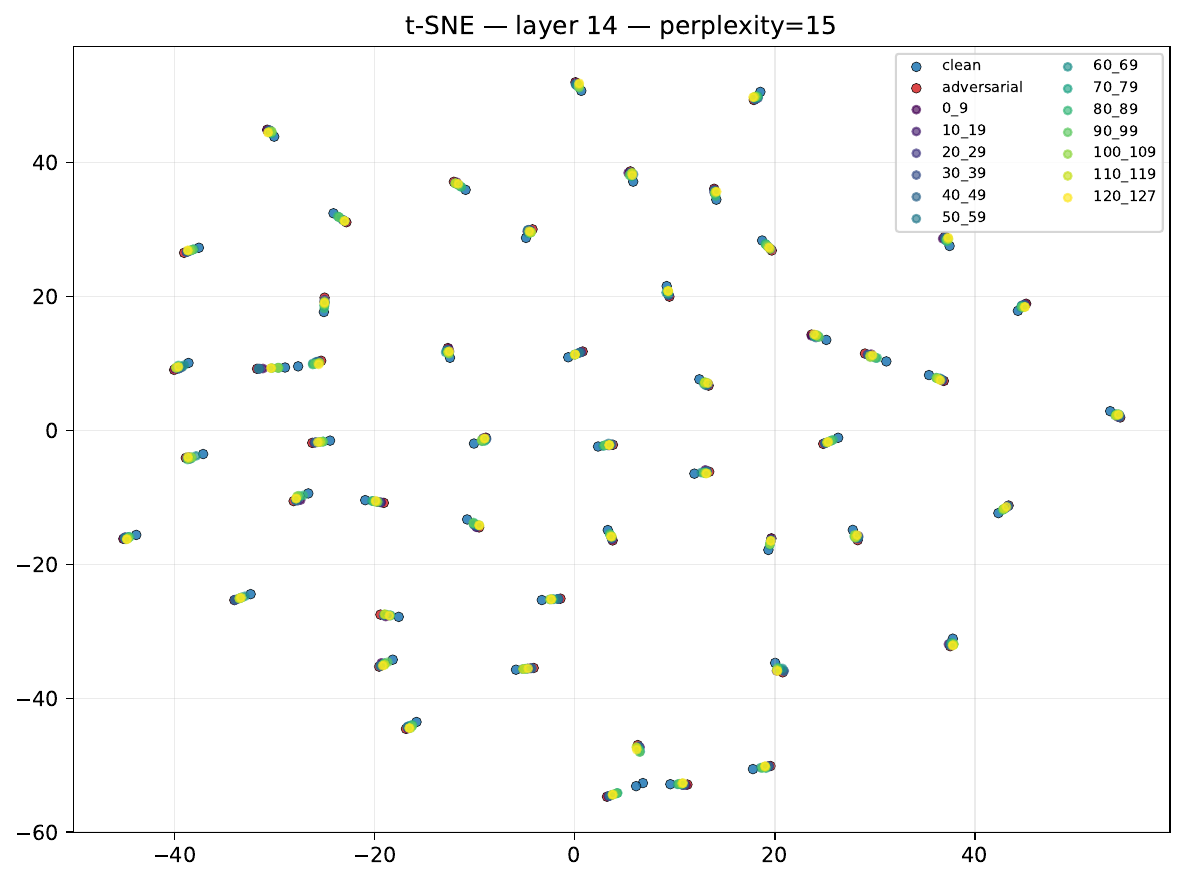}
  \caption{LLaMA-Omni at layer $14$, $n{=}50$. \textbf{Left:} PCA. \textbf{Right:} t-SNE (perplexity $15$). As on Qwen2-Audio (Figure~\ref{fig:pca_tsne_qwen2_l30}), heavy condition overlap confirms a structured high-dimensional displacement rather than a 2D cluster.}
  \label{fig:pca_tsne_omni_l14}
\end{figure}

\vspace{\SubBeforeNeg}
\subsection{Supplemental Analyses and Diagnostics}
\vspace{\SubAfterNeg}
\label{sec:appendix_results}

\vspace{\ParaVspace}
\paragraph{Activation-patching gate and scale detail.}
Before each cross-patch, a self-patch gate checks that replacing a layer's activation with itself reproduces the original generation exactly. The reference activations are collected from the generation-regime forward, following the procedure described by Zhang and Nanda~\citep{zhang2024bestpractices}. The self-patch identity holds at all 312 checks ($12$ samples $\times$ $13$ layers $\times$ $2$ modes). Injected and incumbent spans differ in norm (mean $|\mathrm{ratio}-1|=0.10/0.20/0.37/0.71$ at layers $31/30/28/20$, reaching about $2.6\times$ at layer 20); the norm-matched mode addresses this RMSNorm scale difference. Raw-overwrite and norm-matched results agree (Section~\ref{sec:res_causal}).\label{sec:app_activation}

\vspace{\ParaVspace}
\noindent\textbf{Per-layer restoration and power.} We inject the full adversarial audio-span representation into a band-occluded stream at one layer, using raw-overwrite and norm-matched modes~\citep{zhang2024bestpractices}. Under the pre-specified localization criterion, restoration should increase with depth while the control remains flat. Instead, restoration decreases: it is $0.83$ at the projector output, $0.67$--$0.75$ at layer 8, $0.33$ at layer 12, $0.08$ at layer 16, and zero at layers 18--30. The control band follows a similar pattern, so the early restoration is not specific to the necessary band. Patching the full adversarial span into the clean stream also restores the attack only at the decoder entrance. This intervention tests full-span substitution, not the causal role of an individual band. With 12 samples per layer, the 95\% Clopper--Pearson upper bound after zero restorations is 0.26; the design has at least 90\% power to detect restoration of 0.2 or larger. Because downstream positions have already processed the occluded audio at lower layers, single-layer patching cannot test effects that emerge cumulatively. Cumulative or KV-cache patching is left for future work (Section~\ref{sec:limitations}).

\vspace{\SubBeforeNeg}
\subsubsection{Per-Band Behavioral Detail}
\vspace{\SubAfterNeg}
\label{sec:app_repr_detail}

\begin{figure}[t]
  \centering
  \includegraphics[width=\linewidth]{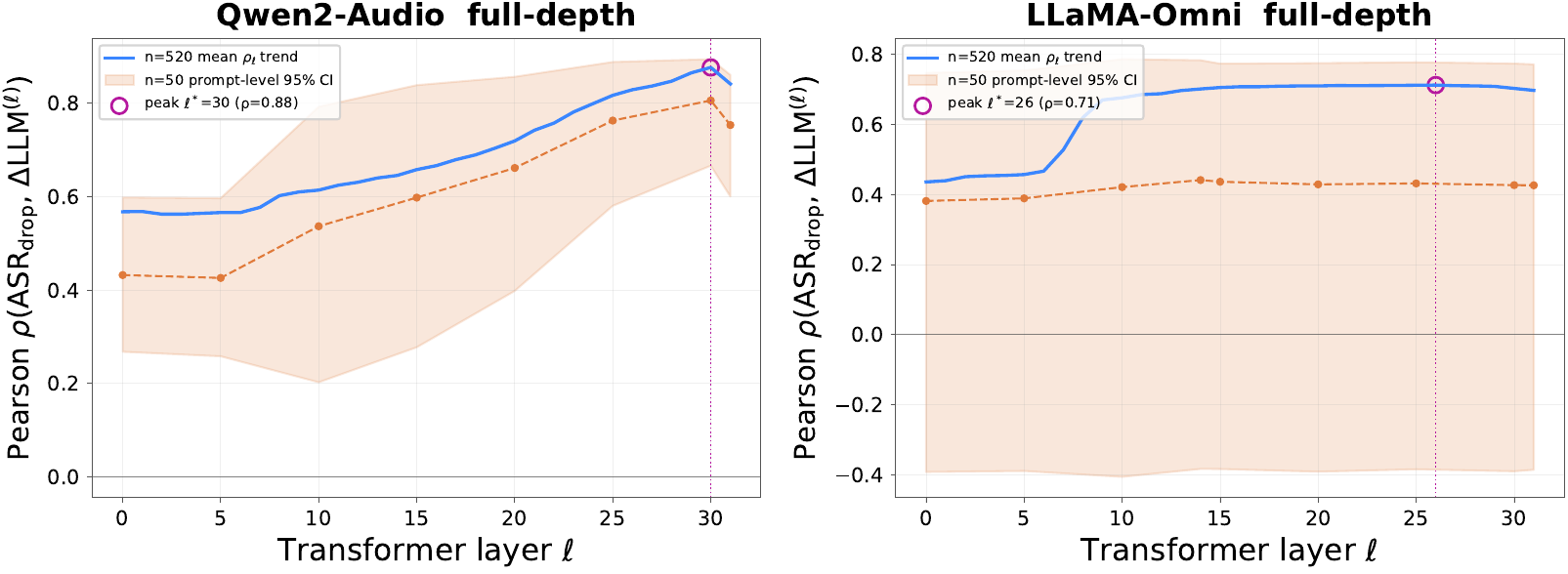}
  \caption{Raw full-depth correlation curve. The blue line shows per-band Pearson correlation $\rho_\ell=\mathrm{corr}(\mathrm{ASR\_Drop},\Delta_{\mathrm{llm}}^{(\ell)})$ from the $n{=}520$ mean divergences across 32 layers. The orange region shows prompt-bootstrap 95\% CIs at layers $\{0,5,10,15,20,25,30,31\}$. On Qwen2-Audio, the point estimate rises from $\rho_0\approx0.568$ at the projector output to $\rho_{30}=0.877$; its raw curve does not reject the phase-randomized surrogate null ($p=0.183$). On LLaMA-Omni, the late-decoder slope is near zero and per-band effects are imprecise. This figure is descriptive; the adjusted analysis appears in Table~\ref{tab:rho_estimators}. The marked layers are visual references, not critical-layer estimates.}
  \label{fig:r12_full_depth}
\end{figure}

\begin{figure}[t]
  \centering
  \includegraphics[width=\linewidth]{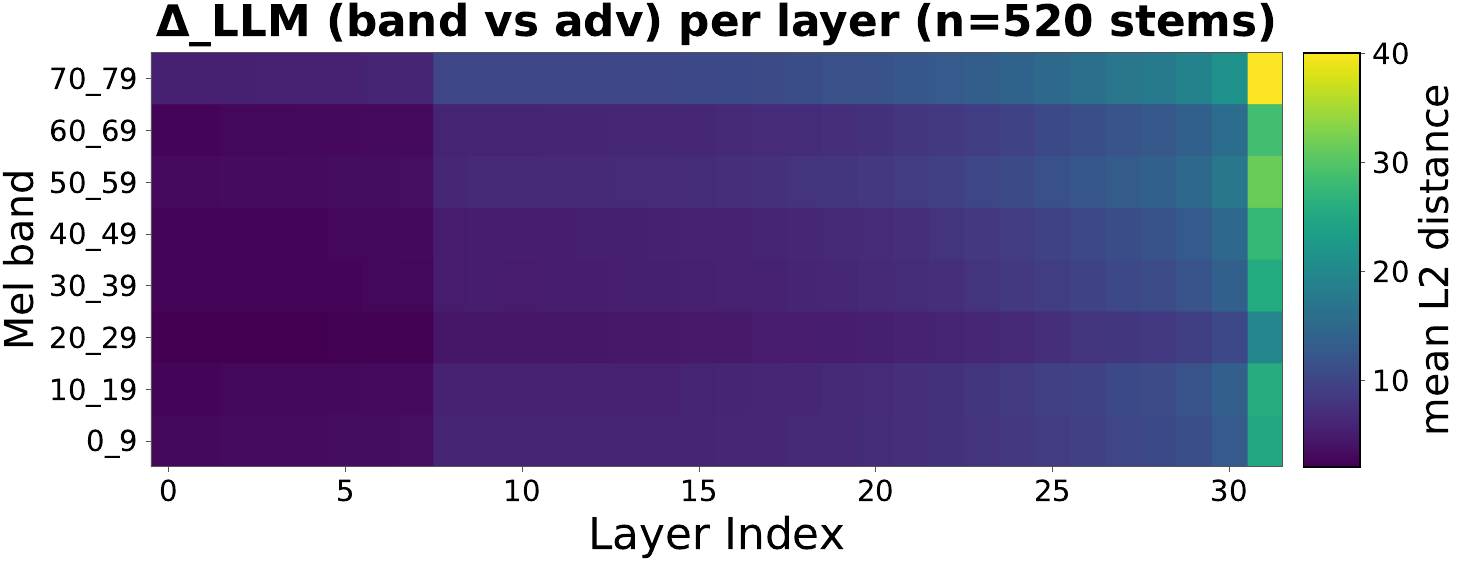}
  \caption{Band-by-layer audio-span divergence on Qwen2-Audio ($32$ decoder layers, $K{=}8$ analysis bands, $n{=}520$). The heatmap shows raw divergence across conditions. The prompt- and energy-adjusted association with band necessity is summarized in Table~\ref{tab:rho_estimators}; the analysis does not identify a causal layer.}
  \label{fig:qwen2_rho}
\end{figure}

\vspace{\ParaVspace}
Figure~\ref{fig:r12_full_depth} shows the pooled-audio-span curve and its prompt-bootstrap band used in Section~\ref{sec:res_repr}. The last-audio-frame correlation stays lower and nearly flat ($\rho^{\mathrm{last}}\approx0.45$--$0.49$); Figure~\ref{fig:qwen2_rho} shows divergence by band and layer.

\begin{table}[b]
  \centering
  \small
  \setlength{\tabcolsep}{4pt}
  \begin{tabular}{@{}lcccl@{}}
    \toprule
    Estimator & $n$ & Initial value & Layer 30 & Contrast \\
    \midrule
    Pooled raw, $K{=}8$ & 520 & 0.568 & 0.877 & $\Delta\rho=+0.309$ \\
    Prompt-bootstrap raw & 520 & 0.564 & 0.872 & $\Delta\rho=+0.307$ $[0.271,0.342]$ \\
    Disattenuated raw & 520 & 0.568 & 0.878 & $\Delta\rho=+0.310$ \\
    Max-pooled span & 520 & 0.665 & 0.732 & $\Delta\rho=+0.067$ (peak 0.84) \\
    Cosine, norm-normalized & 520 & 0.501 & 0.835 & $\Delta\rho=+0.334$ \\
    Continuous judge score & 520 & 0.58 & 0.88 & $\Delta\rho=+0.30$ \\
    Raw, equal-energy $K{=}16$ & 50 & 0.79 & 0.92 & $\Delta\rho=+0.13$ \\
    Adjusted mixed-effects $\gamma_\ell$ & 399 & $\AGammaZero$ & $\AGammaThirty$ & $\gamma_{30}-\gamma_0=+0.293$ $[0.11,0.51]$ \\
    \bottomrule
  \end{tabular}
  \vspace{-0.3cm}
  \caption{Raw depth-alignment estimates and the prompt- and energy-adjusted mixed-effects model. Initial and layer-30 values are correlations $\rho_\ell$ except in the final row, which reports coefficients $\gamma_\ell$. The prompt-bootstrap interval for the raw $K{=}8$ increase excludes zero, but the leave-one-band-out spread is $0.172$, showing sensitivity to this partition; the phase-randomized test of the raw late-decoder slope gives $p=0.183$. The adjusted model uses the 399 prompts jailbroken by the full adversarial input and per-prompt, per-band energy. Its layer-30 coefficient is $+0.93$ (95\% CI $[0.70,1.19]$), and the pre-specified layer-30-minus-projector contrast is $+0.293$ (95\% CI $[0.11,0.51]$). The behavior-irrelevant controls show no corresponding depth modulation, but are defined over the same coarse frequency partition. The $K{=}16$ raw-correlation estimate uses $n{=}50$; its behavioral sweep covers $n{=}520$, but per-stem $K{=}16$ hidden states were not collected.}
  \label{tab:rho_estimators}
\end{table}

\begin{figure}[t]
  \centering
  \includegraphics[width=\linewidth]{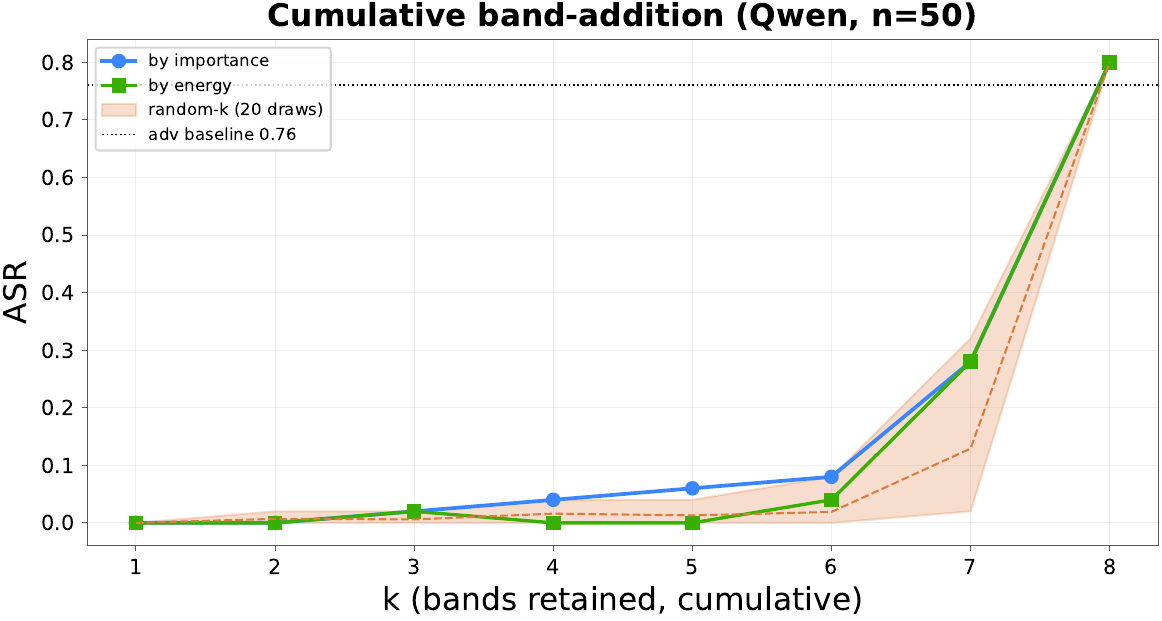}
  \caption{Cumulative band-addition on Qwen2-Audio ($n{=}50$). ASR as a function of $k$ retained bands, ordered by mask-sweep importance and by retained energy, against an energy-matched random-$k$ baseline (orange band, $20$ draws). Retaining the least-important bands first leaves ASR at the floor through $k{=}6$ (importance path $0.00,0.00,0.02,0.04,0.06,0.08$), rises to $0.28$ at $k{=}7$, and reaches the full adversarial ASR only when all bands are retained ($k{=}8$, $0.80$). The importance- and energy-ordered paths are closely matched, so \emph{retention order is immaterial}; the energy-matched \emph{random} baseline stays near zero until $k{=}7$--$8$, which is the control's purpose: it shows the sufficiency signal is carried by bandwidth+energy coverage, not by which specific bands are retained.}
  \label{fig:r2_cumulative}
\end{figure}

\begin{table}[t]
  \centering
  \small
  \begin{tabular}{lccc}
    \toprule
    condition & $\mathrm{ASR}_{\mathrm{primary}}$ & $\mathrm{ASR}_{\mathrm{ref}}$ &
      agreement($\kappa$) \\
    \midrule
    adversarial (no mask) & 0.76 & 0.64 & 0.80 ($0.53$) \\
    $b_1=[0,9]$    & 0.32 & 0.28 & 0.92 ($0.81$) \\
    $b_2=[10,19]$  & 0.18 & 0.14 & 0.96 ($0.85$) \\
    $b_3=[20,29]$  & 0.28 & 0.24 & 0.92 ($0.79$) \\
    $b_4=[30,39]$  & 0.18 & 0.14 & 0.96 ($0.85$) \\
    $b_5=[40,49]$  & 0.04 & 0.04 & 1.00 ($1.00$) \\
    $b_6=[50,59]$  & 0.04 & 0.04 & 1.00 ($1.00$) \\
    $b_7=[60,69]$  & 0.08 & 0.08 & 1.00 ($1.00$) \\
    $b_8=[70,79]$  & 0.02 & 0.02 & 1.00 ($1.00$) \\
    \bottomrule
  \end{tabular}
\caption{Agreement between the primary AdvWave judge (GPT-4o-mini) and a stricter reference judge (GPT-5.6-sol) on the $K{=}8$ mask sweep and adversarial condition ($n{=}50$ per condition, Qwen2-Audio, $\varepsilon{=}0.01$). These re-scored ASRs come from an $n{=}50$ sensitivity run, not the headline float32 results in Table~\ref{tab:recon_path}. The stricter judge labels fewer borderline adversarial responses as jailbreaks, but the judges preserve the band ordering (Section~\ref{sec:limitations}). We use GPT-4o-mini for the primary evaluation and GPT-5.6-sol only as a sensitivity check.}
  \label{tab:judge_agreement}
\end{table}

\vspace{\SubBeforeNeg}
\subsubsection{Judge-Misclassification Correction (Rogan-Gladen)}
\vspace{\SubAfterNeg}
\label{sec:app_judge}

\vspace{\ParaVspace}
We validate the primary judge on the condition-blind sample described in Section~\ref{sec:meth_inference}. Among the 120 adversarial responses, the judge labels $P_j{=}0.733$ as jailbreaks ($88/120$); the annotator labels $0.825$ as jailbreaks ($99/120$). The estimated conditional false-positive rate is $FPR{=}0.095$ ($2/21$ human negatives), and the false-negative rate is $FNR{=}0.131$ ($13/99$ human positives). The Rogan--Gladen estimate $(P_j-FPR)/(1-FPR-FNR)$ applied to the $n{=}520$ judged adversarial ASR of $0.767$ is $0.868\approx0.87$. Uncertainty in the validation rates gives an approximate range of $0.83$--$0.95$ (Section~\ref{sec:limitations}). These estimates apply to the adversarial condition only; they do not correct masked-condition ASRs.

\vspace{\ParaVspace}
Table~\ref{tab:recon_path} reports the primary $K{=}8$ contiguous mask sweep. The perturbation is transformed and reconstructed in float32 using an STFT with $n_{\mathrm{fft}}{=}400$ and hop length $160$ at $16$\,kHz, without clipping or int16 conversion. A direct waveform comparison gives relative RMSE $1.76{\times}10^{-5}$ and signal-to-quantization-noise ratio of about $70.5$\,dB; quantization is small relative to the signal.

\begin{table}[t]
  \centering
  \small
  \setlength{\tabcolsep}{5pt}
  \begin{tabular}{@{}lc@{}}\toprule
    Mel-band & ASR \\
    $(\bm{\delta}\ \mathrm{bins})$ & (float32) \\
    \midrule
    0--9    & 0.24 \\
    10--19  & 0.17 \\
    20--29  & 0.29 \\
    30--39  & 0.14 \\
    40--49  & 0.11 \\
    50--59  & 0.07 \\
    60--69  & 0.09 \\
    70--79  & 0.05 \\
    \bottomrule
  \end{tabular}
  \vspace{-0.05cm}
\caption{Per-band jailbreak ASR for the primary $K{=}8$ contiguous mask sweep ($n{=}520$), using float32 STFT-mask-ISTFT reconstruction without clipping. Direct waveform comparison gives a relative RMSE of $1.76{\times}10^{-5}$ and signal-to-quantization-noise ratio of about 70.5~dB.}
  \label{tab:recon_path}
\end{table}

\vspace{\SubBeforeNeg}
\subsubsection{Energy Diagnostics}
\vspace{\SubAfterNeg}
\label{sec:app_energy_diag}

\vspace{\ParaVspace}
Table~\ref{tab:protocol_energy} reports each band's share of total $\bm\delta$ energy, its Hz edges, the $n{=}520$ $K{=}8$ mask ASR drop, and the round-trip occlusion residual (the fraction of in-band energy that re-enters the supposedly removed band), the data underlying the energy-confounded reading in \S\ref{sec:res_fidelity}.

\begin{table}[t]
  \centering
  \small
  \setlength{\tabcolsep}{3.5pt}
  \begin{tabular}{@{}cc cccc cc@{}}
    \toprule
    & & \multicolumn{4}{c}{Qwen2-Audio (80-bin analysis basis)} & \multicolumn{2}{c}{LLaMA-Omni (128-bin analysis basis)} \\
    \cmidrule(lr){3-6}\cmidrule(lr){7-8}
    Band & Hz $[f_{\mathrm{low}}, f_{\mathrm{high}}]$ & En. & Res. & Drop & {Hz} & En. \\
    \midrule
    $[0,9]$    &  0--344    & 0.10 & 0.014 & 0.53 &  0--214    & 0.056 \\
    $[10,19]$  & 382--725  & 0.089 & 0.030 & 0.60 & 238--451  & 0.047 \\
    $[20,29]$  & 764--1117 & 0.072 & 0.035 & 0.48 & 475--689  & 0.051 \\
    $[30,39]$  & 1162--1656& 0.101 & 0.022 & 0.63 & 713--926  & 0.046 \\
    $[40,49]$  & 1722--2455& 0.105 & 0.017 & 0.66 & 950--1184 & 0.046 \\
    $[50,59]$  & 2554--3640& 0.146 & 0.010 & 0.70 & 1213--1513& 0.056 \\
    $[60,69]$  & 3786--5396& 0.147 & 0.007 & 0.68 & 1550--1932& 0.073 \\
    $[70,79]$  & 5613--8000& 0.242 & 0.003 & 0.72 & 1980--2469& 0.071 \\
    $[80,89]$  & --        & --    & --     & --  & 2530--3154& 0.078 \\
    $[90,99]$  & --        & --    & --     & --  & 3232--4029& 0.087 \\
    $[100,109]$& --        & --    & --     & --  & 4129--5148& 0.084 \\
    $[110,119]$& --        & --    & --     & --  & 5275--6576& 0.084 \\
    $[120,127]$& --        & --    & --     & --  & 6739--8000& 0.221 \\
    \bottomrule
  \end{tabular}
  \vspace{-0.05cm}
  \caption{Band-level perturbation-energy share (En.), Hz boundaries, reconstruction residual (Res.), and mask-sweep ASR drop for Qwen2-Audio, with energy share and Hz boundaries also shown for LLaMA-Omni. Boundaries use the Slaney mel scale over 0--8 kHz; we report frequencies in Hz so the two analysis bases can be compared (Section~\ref{sec:setup}). The residual is the fraction of energy that re-enters the masked band after STFT masking, inverse STFT, and re-analysis. All checked residuals are below 5\%, and neighboring-band mel-filter leakage is below 1.5\% of the in-band attenuation. The residual diagnostic uses complete 10-bin groups, so Omni's final 8-bin group is not measured. ASR drop is $0.767$ minus the masked ASR in Table~\ref{tab:recon_path}.}
  \label{tab:protocol_energy}
\end{table}

\vspace{\SubBeforeNeg}
\subsubsection{LLaMA-Omni Behavior}
\vspace{\SubAfterNeg}
\label{sec:app_omni}

\begin{figure}[t]
  \centering
  \includegraphics[width=\linewidth]{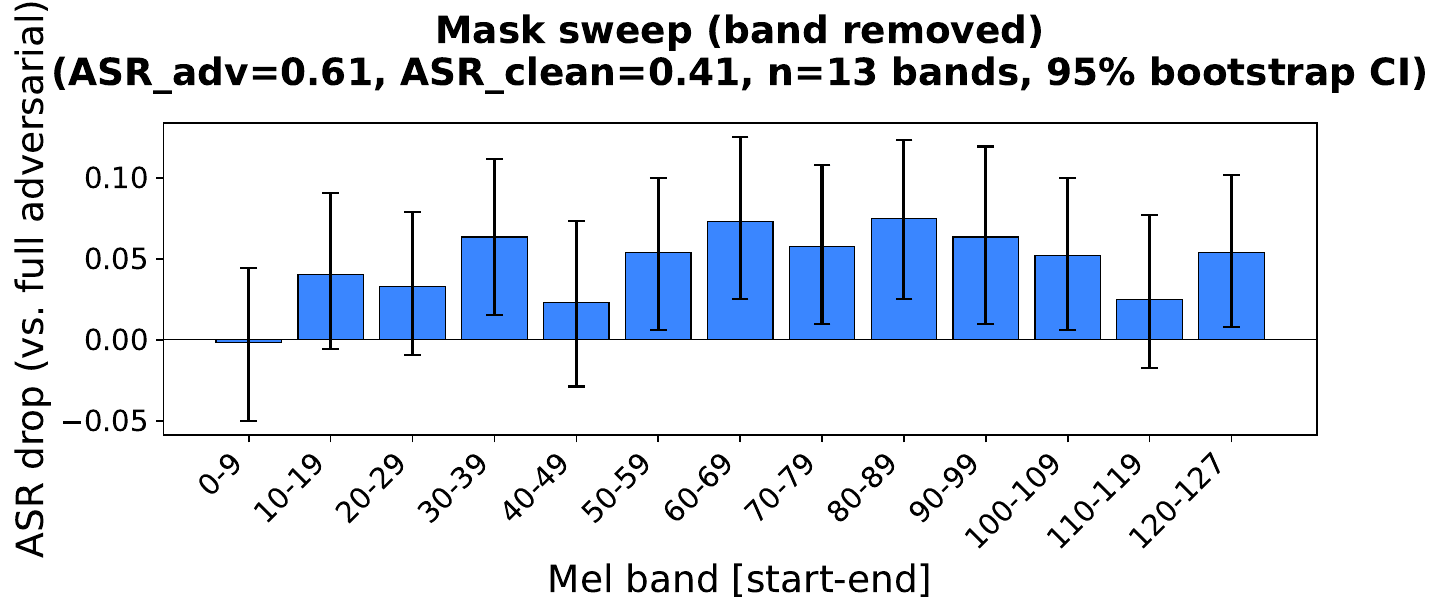}
  \caption{LLaMA-Omni null arm ($128$-bin analysis basis, $K{=}13$). We read this model as a null, not as a power study, because its small margin and judge noise prevent reliable per-band comparisons (Section~\ref{sec:res_supplemental}). The plot shows mask-sweep ASR drops relative to the adversarial baseline of $0.61$ ($n{=}520$); every 95\% bootstrap CI overlaps zero. The raw depth estimates are summarized in Table~\ref{tab:cross_model_summary}. Omni's 128-bin analysis basis covers different frequencies from Qwen's 80-bin basis; Table~\ref{tab:protocol_energy} reports the corresponding Hz boundaries.}
  \label{fig:omni_asr_and_rho}
\end{figure}

\vspace{\ParaVspace}
The Omni ASR and depth-alignment sweeps are reported in Figure~\ref{fig:omni_asr_and_rho}, and the cross-architecture summary (mask/keep band drops and the depth-slope row) in Table~\ref{tab:cross_model_summary}.

\begin{table}[t]
  \centering
  \footnotesize
  \setlength{\tabcolsep}{4pt}
  \renewcommand{\arraystretch}{1.15}
  \begin{tabular*}{\linewidth}{@{\extracolsep{\fill}} lcc @{}}
    \toprule
    \textbf{Metric} & \textbf{Qwen2-Audio} & \textbf{LLaMA-Omni} \\
    \midrule
    Evaluated prompts $n$            & 520          & 520 \\
    Analysis bands $K$               & 8            & 13 \\
    \addlinespace
    ASR, clean                       & 0.008        & 0.41 \\
    ASR, adversarial                 & 0.77         & 0.61 \\
    \addlinespace
    Top mask band (basis index)      & $[70, 79]$   & $[80, 89]$ \\
    Top mask band (frequency)        & 5.6--8.0 kHz & 2.5--3.2 kHz \\
    \text{ASR\_Drop} at top band     & 0.72         & 0.08 \\
    \addlinespace
    Late-decoder $\rho_\ell$ slope   & rising       & $\approx$ flat \\
    \bottomrule
  \end{tabular*}
  \vspace{-0.4cm}
  \caption{Cross-architecture comparison on the $n{=}520$ corpus. LLaMA-Omni is interpreted as a null arm because its clean ASR leaves too little margin for reliable per-band comparisons (Section~\ref{sec:res_supplemental}). ASR drops are measured relative to each model's own adversarial baseline. The depth row reports the late-decoder slope of the raw $\rho_\ell$ curve, not a selected peak layer. We omit keep-only results here because they are floor-saturated on Qwen (all retained-band ASRs are $\leq0.02$) and return toward Omni's clean ASR; the cumulative band-addition curve is the more informative sufficiency analysis (Section~\ref{sec:res_asr}).}
  \label{tab:cross_model_summary}
\end{table}

\vspace{\SubBeforeNeg}
\subsubsection{Detector Depth Curves}
\vspace{\SubAfterNeg}
\label{sec:app_detector}

\begin{figure}[t]
  \centering
  \includegraphics[width=\linewidth, trim = 0 15 0 10]{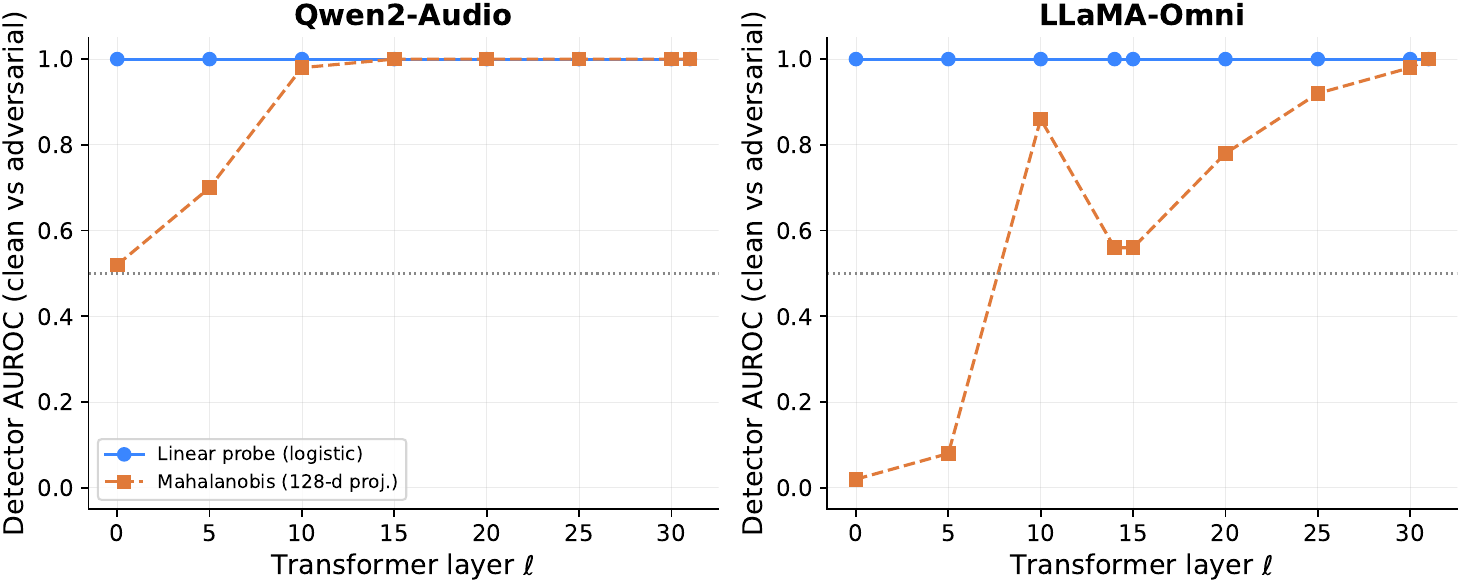}
  \caption{Clean-versus-adversarial detector AUROC as a function of decoder depth, on the $n{=}50$ pooled audio-span hidden states. A linear probe (logistic regression, held out by prompt) separates clean from adversarial representations with AUROC $1.0$ at every sampled layer on both ALMs; a Mahalanobis detector on a $128$-dim random projection rises from chance at the input layers to near-perfect at depth. The jailbreak displacement is therefore large enough to be linearly detectable in the high-dimensional residual stream even though the first two PCA/t-SNE axes show no cluster.}
  \label{fig:b6_detector_auroc}
\end{figure}

\vspace{\ParaVspace}
Figure~\ref{fig:b6_detector_auroc} reports the linear-probe and Mahalanobis detector AUROC by decoder depth on both ALMs. The input-versus-residual-stream comparison (adversarial-versus-benign-perturbed, plus the untrained-epsilon column) is in Table~\ref{tab:detector_rebaseline}, discussed in \S\ref{sec:res_supplemental}.

\begin{table}[t]
  \centering
  \small
  \setlength{\tabcolsep}{4pt}
  \begin{tabular}{@{}lcccc@{}}
    \toprule
    Detector & & \multicolumn{2}{c}{AUROC} & \multicolumn{1}{c}{Untrained-$\epsilon$ AUROC} \\
    & & clean vs. adv & adv vs. benign-perturbed & (0.005\,/\,0.02) \\
    \midrule
    Input space, high-freq. energy ratio &  & 0.92 & 0.64 & --- \\
    Input space, spectral flatness       &  & 1.00 & 0.57 & --- \\
    \midrule
    Depth linear probe (layer 30)        &  & 1.00 & 1.00 & 1.00\,/\,1.00 \\
    Input space, log-mel logistic ($d{=}80$) & & 1.00 & 0.999 & --- \\
    \bottomrule
  \end{tabular}
  \vspace{-0.2cm}
  \caption{Input-space and residual-stream detector AUROCs on the Qwen2-Audio subset ($n{=}50$). The first column separates clean from adversarial audio. The second tests adversarial audio against benignly corrupted speech, using additive noise at SNR $15$ or $25$ dB and MP3 or Opus compression at 64 kbps. Unsupervised input features perform well on the clean-versus-adversarial task but decline with benign corruption. The depth probe remains at 1.0; however, a supervised logistic-regression classifier on the victim's log-mel input features reaches 1.0 and 0.999 on the two tasks, respectively. Thus, this experiment shows no measurable AUROC advantage for the depth probe over a comparable supervised input baseline. The final column evaluates a depth probe trained only on $\epsilon=0.01$ samples against attacks optimized at $\epsilon=0.005$ and $0.02$ (Section~\ref{sec:res_supplemental}). AUROC remains 1.00, indicating that the probe detects the perturbation family, not successful jailbreak behavior at a particular budget.}
  \label{tab:detector_rebaseline}
\end{table}

\vspace{\SubBeforeNeg}
\subsubsection{Cross-Model Transfer}
\vspace{\SubAfterNeg}
\label{sec:app_transfer}

\begin{figure}[!t]
  \centering
  \includegraphics[width=\textwidth, trim=5 10 0 20]{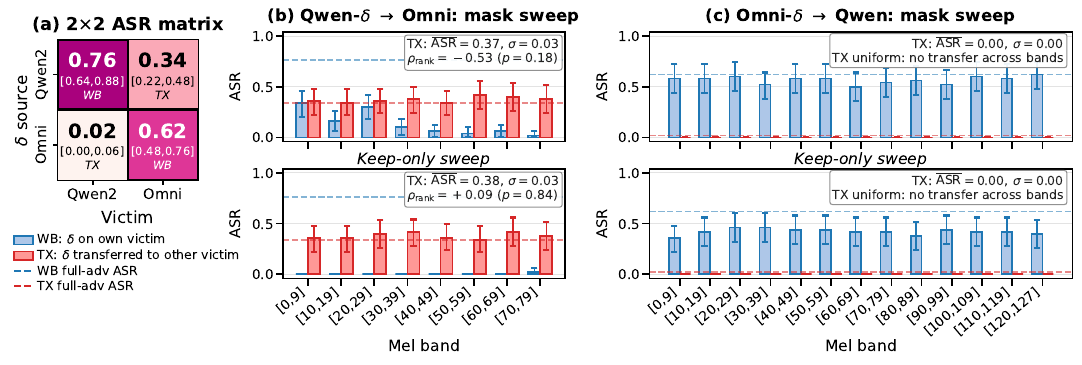}
  \vspace{-0.3cm}
  \caption{Cross-model AdvWave-P transferability on the $n{=}50$ matched subset. \textbf{(a)} $2{\times}2$ ASR matrix; diagonal cells (\emph{WB}) are white-box, off-diagonal (\emph{TX}) are direct black-box transfer. \textbf{(b,\,c)} per-band mask (top) and keep-only (bottom) sweeps in both directions: blue bars are the source-victim white-box ASR, red bars are the cross-victim transfer ASR; dashed lines mark the full-adversarial WB and TX ASRs. Inline boxes report transfer mean$\pm$$\sigma$ across bands and Spearman $\rho_\text{rank}$ between the WB and TX ASR-drop rankings. Brackets in (a) are bootstrap $95\%$ CIs over $1000$ resamples.}
  \label{fig:transferability_matrix}
\end{figure}

\vspace{\ParaVspace}
Figure~\ref{fig:transferability_matrix} reports the $2{\times}2$ ASR matrix and the per-band sweep across each cross-victim. The transfer result is a two-directional null, summarized in \S\ref{sec:res_supplemental}.

\vspace{\SubBeforeNeg}
\subsection{Clean-Speech Renderer}
\vspace{\SubAfterNeg}
\label{sec:appendix_renderer}
\vspace{\ParaVspace}
We render the clean AdvBench waveforms with OpenAI's \texttt{tts-1-hd} API using the \texttt{alloy} voice at speed $1.0$. The service returns 24\,kHz audio in an MP3 container, stored with a .wav filename; we resample it to 16\,kHz mono for both victims. Each clean waveform is decoded once, and the same 16\,kHz samples are used during perturbation optimization and all later inference. Qwen2-Audio and LLaMA-Omni therefore receive byte-identical clean inputs, enabling a direct comparison of their clean-condition results (Section~\ref{sec:res_supplemental}).

\vspace{\SubBeforeNeg}
\subsection{Artifact Availability and Reproducibility}
\vspace{\SubAfterNeg}
\label{sec:appendix_artifacts}

\vspace{\ParaVspace}
The accompanying repository will provide analysis and optimization code, random seeds, attack hyperparameters, renderer specifications, prompt identifiers, model responses, and per-condition judge scores needed to re-derive the reported results from the public prompt corpora. It will not distribute optimized adversarial waveforms or perturbations. The primary AdvBench audio was generated with OpenAI \texttt{tts-1-hd}, voice \texttt{alloy}, then resampled to 16\,kHz mono. The three additional corpora used \texttt{edge-tts} with voice \texttt{en-US-AriaNeural} (Appendix~\ref{sec:appendix_renderer} and Section~\ref{sec:app_second_corpus}). Because the judges are provider-hosted models without archived immutable snapshots, released responses and scores, rather than future re-queries to those services, are the reference for reproducing the labels.

\vspace{\SubBeforeNeg}
\subsection{Gradient Band-Attribution Baseline}
\vspace{\SubAfterNeg}
\label{sec:app_grad_attr}

\vspace{\ParaVspace}
Occlusion requires $K{+}2$ forward passes per sample, so gradient attribution offers a cheaper ranking check. On 50 saved $\varepsilon=0.01$ adversarial waveforms, we backpropagate AdvWave-P's next-token cross-entropy objective to $\bm\delta$ and sum squared gradients within each analysis band in the STFT domain. Without re-optimization or generation, this gradient ranking correlates with the occlusion ranking (Spearman $\rho=0.97$, $p<0.001$; Pearson $0.84$). Energy share alone also recovers much of the ranking at $n=520$ (Spearman $\rho=0.95$). These baselines show that occlusion is not an independent source of the ranking; its added value is the set of controls for bandwidth, energy, and removal structure. Using gradient saliency as the behavioral target also produces a raw depth curve that rises from $\rho_0=0.70$ to $\rho_{30}=0.94$ (late-decoder slope $+0.0102$ per layer). This curve tracks band geometry and is not evidence for a behavior-specific depth effect.

\vspace{\SubBeforeNeg}
\subsection{Decoding and Optimization Seed Stability}
\vspace{\SubAfterNeg}
\label{sec:app_seed_stability}

\vspace{\ParaVspace}
\paragraph{Decoding seeds.} The main sweep uses greedy decoding. To check sensitivity to decoding, we re-evaluate the adversarial condition and the highest- and lowest-importance masked bands with stochastic decoding ($\mathrm{do\_sample}{=}\mathrm{true}$, $t{=}1.0$, three seeds, $n{=}520$, cold start). Adversarial ASR varies from $0.56$ to $0.75$ (mean $0.67$), below the greedy result of $0.77$. The tested band contrast is consistent across seeds: masking the top band $[70,79]$ leaves ASR at $0.04$--$0.11$ (mean $0.07$), while masking the low band $[0,9]$ leaves ASR at $0.18$--$0.31$ (mean $0.24$). These two conditions do not establish a full-band ranking under stochastic decoding. The variation in adversarial ASR is larger than expected from binomial noise alone (per-seed standard error at $n{=}520$ is at most $0.022$); its source remains unresolved. We therefore treat this check as evidence about these two masks, not as a replication of the complete ranking.

\vspace{\ParaBeforeNeg}
\paragraph{Optimization seeds.} We also tested whether the perturbation's energy profile depends on its random initialization. On 20 prompts, we re-optimized AdvWave-P at $\varepsilon=0.01$ for 3,000 Adam iterations from three independent initializations. Every prompt placed its largest energy share in the same band, $b_8=[70,79]$ (20/20 prompts). Across seeds, the mean pairwise Spearman correlation of per-band energy shares was $0.97$ (minimum $0.88$); 19 of 20 prompts had all three pairwise correlations at or above $0.9$. The largest band's energy share averaged $0.21$ (SD $0.015$), and every perturbation respected the $L_\infty$ bound. This indicates that the top-energy band is stable across these optimization seeds. It does not establish frequency-specific necessity, since the band's larger frequency width also contributes to its energy share (Section~\ref{sec:res_asr}).

\vspace{\SubBeforeNeg}
\subsection{An Arbitrary Single Notch as a Defense Hypothesis}
\vspace{\SubAfterNeg}
\label{sec:app_defense}

\vspace{\ParaVspace}
A single band-stop notch can substantially reduce attack success even when applied to a low-frequency band. Masking $b_1{=}[0,9]$, which contains about $10\%$ of the perturbation energy and is the second-least-important band in the $n{=}520$ sweep, reduces ASR from $0.77$ to $0.24$ ($n{=}520$); the $n{=}50$ subset drops from $0.76$ to $0.34$. This is a property of the tested perturbation, not a validated defense. A practical filter would have to act on the mixture $x{+}\bm\delta$, and a broad low-frequency notch could also remove speech fundamentals. We did not measure speech utility and therefore do not recommend this intervention.

\vspace{\SubBeforeNeg}
\subsection{Resolution Ladder Across \texorpdfstring{$K$}{K}}
\vspace{\SubAfterNeg}
\label{sec:app_ladder}

\vspace{\ParaVspace}
This appendix reports exploratory raw correlations across partition resolutions. They are descriptive sensitivity checks and do not replace the prompt- and energy-adjusted model in Section~\ref{sec:res_repr}. Energy-adjusted partial-correlation estimates are omitted because the near-collinearity produced invalid values outside $[-1,1]$.

\vspace{\ParaVspace}
We test whether the raw depth increment changes with partition resolution by repeating the analysis at $K{=}4$ and $K{=}20$, alongside the $K{=}8$ and $K{=}16$ results. At $K{=}4$, band-mask drops saturate, leaving too little across-band variation for a reliable depth estimate. At the resolutions where band effects vary, the raw increments are $+0.36$ ($K{=}8$), $+0.13$ ($K{=}16$), and $+0.094$ ($K{=}20$), each at $n{=}50$; the full-corpus $K{=}8$ estimate is $+0.309$. These point estimates depend on the partition and are descriptive, not evidence for a resolution-independent depth effect. The primary $K{=}8$, $n{=}520$ trend does not reject the autocorrelation-preserving null (Section~\ref{sec:res_repr}). Energy-adjusted partial correlations at $K{=}8$ are degenerate, with estimates outside $[-1,1]$ because band energy and representational divergence are near-collinear; we therefore omit energy-conditioned depth increments rather than interpret them.

\vspace{\SubBeforeNeg}
\subsection{Audit of Named Prior Band-Selectivity Claims}
\vspace{\SubAfterNeg}
\label{sec:app_audit}

\vspace{\ParaVspace}
GRM ranks mel bands by an attack-to-utility ratio and perturbs a selected subset~\citep{wang2026grm}. ALMGuard selects mel bins for applying a defensive universal perturbation, balancing jailbreak mitigation with speech-recognition performance~\citep{jin2025almguard}. Neither selection rule is, by itself, a claim about which components of an attacker's perturbation are necessary. We applied the selected supports to AdvWave-P as an exploratory audit using random controls. The ALMGuard comparison used equal-energy suppression; the GRM comparison used random supports matched in bandwidth. Both pilots were inconclusive and neither evaluates the methods' original objectives.

\vspace{\ParaVspace}
\textbf{ALMGuard (equal-energy suppression, $n{=}520$).} We map ALMGuard's released $128$-mel selection mask for Qwen2-Audio to the $201$-bin STFT analysis basis. We suppress the selected coefficients so that the removed perturbation energy is approximately $10\%$, then compare with 20 random masks matched on removed energy and bandwidth. The ALMGuard condition gives ASR $\AlmAsr$; random masks have mean $\AlmRandMean$ and empirical range $[0.150,0.187]$. Its result falls inside this range, so the pilot does not resolve a difference. This comparison tests the selected mask on AdvWave-P; it is not an evaluation of ALMGuard's defensive universal perturbation.

\vspace{\ParaVspace}
\textbf{GRM (band-restricted re-optimization, $n{=}20$).} We reproduce GRM's selection rule, not its original perturbation, by restricting a newly optimized AdvWave-P perturbation to its selected 109 of 201 STFT bins. Cold-start inference on 20 jailbroken JailbreakBench prompts, scored with GPT-4o-mini, gives ASR $\GrmAsr$ ($15/20$). Twenty random supports matched in bandwidth reach mean ASR $\GrmRandMean$ (range $[\GrmRandLo,\GrmRandHi]$; bootstrap 95\% CI for the mean $[\GrmRandBoot,\GrmRandBootHi]$). The GRM support lies at the lower edge of the random range. Given this sample size and support construction, the result is inconclusive; the uniformly sampled random supports are scattered and not structurally matched to a contiguous support.

\vspace{\ParaVspace}
Neither pilot resolves whether the selected supports differ from the corresponding random controls. The ALMGuard comparison matches removed energy and bandwidth; the GRM pilot matches bandwidth but uses scattered random supports, unlike the selected support's structure. These exploratory tests do not evaluate the original methods or show that their selections are ineffective. They only test the selected regions on AdvWave-P under the stated conditions.

\vspace{\SubBeforeNeg}
\subsection{Second Corpus: Cross-Corpus Replication}
\vspace{\SubAfterNeg}
\label{sec:app_second_corpus}

\vspace{\ParaVspace}
To examine whether the AdvBench findings extend to other prompt sets, we analyze JailbreakBench (89 deduplicated prompts), MaliciousInstruct (100), and StrongREJECT (276 prompts not in AdvBench). These corpora were rendered with \texttt{edge-tts}, voice \texttt{en-US-AriaNeural}, and attacked at $\varepsilon=0.01$, then evaluated from a cold start. The $K{=}8$ JailbreakBench mask sweep uses all 70 prompts jailbroken by the full adversarial input, without selection on band necessity. For MaliciousInstruct and StrongREJECT, the available analyses use the 50 highest-necessity prompts selected from the jailbroken pools (93/100 and 190/276, respectively). We follow the primary protocol with Qwen2-Audio, GPT-4o-mini at temperature 0, and float32 reconstruction without clipping; the sample selection differs across these analyses.

\vspace{\ParaVspace}
Table~\ref{tab:second_corpus_mask} reports the $K=8$ mask-sweep drops relative to each sample's adversarial baseline. In the full JailbreakBench pool, masking $70$--$79$ gives the largest drop and masking $20$--$29$ the smallest. The same ordering appears in the necessity-selected MaliciousInstruct and StrongREJECT subsets, but that agreement is conditioned on selection and is not an independent replication. These data do not establish a common ordering at $K=16$.

\begin{table}[t]
\centering\small
\begin{tabular}{lrrrrr}
\toprule
band & JBB (full pool, n=70) & Malicious ($n=50^{\dagger}$) & StrongRE ($n=50^{\dagger}$) & AdvBench (n=520) \\
\midrule
0--9   & $+0.70$ & $+0.82$ & $+0.60$ & $+0.68$ \\
10--19 & $+0.69$ & $+0.76$ & $+0.62$ & $+0.78$ \\
20--29 & $+0.45$ & $+0.60$ & $+0.45$ & $+0.63$ \\
30--39 & $+0.75$ & $+0.74$ & $+0.54$ & $+0.81$ \\
40--49 & $+0.73$ & $+0.80$ & $+0.69$ & $+0.86$ \\
50--59 & $+0.78$ & $+0.87$ & $+0.75$ & $+0.91$ \\
60--69 & $+0.77$ & $+0.88$ & $+0.73$ & $+0.88$ \\
70--79 & $+0.89$ & $+0.96$ & $+0.81$ & $+0.93$ \\
\bottomrule
\end{tabular}
\caption{Normalized $K{=}8$ mask-sweep drop, $1-\mathrm{ASR}(\mathrm{mask})/\mathrm{ASR}_{\mathrm{adv}}$, across corpora. JailbreakBench uses its full jailbroken pool ($n{=}70$ of 89; adversarial ASR $0.914$). $^{\dagger}$MaliciousInstruct and StrongREJECT use the 50 highest-necessity prompts from their jailbroken pools ($93/100$ and $190/276$); those two rankings are selection-conditioned and are shown descriptively, not as independent replication. AdvBench uses $n{=}520$ (adversarial ASR $0.767$).}
\label{tab:second_corpus_mask}
\end{table}

\vspace{\ParaVspace}
On the available 50-prompt samples, cumulative-retention curves stay near zero until roughly the last two bands and then rise toward $0.9$--$1.0$; these curves are selection-conditioned and are not full-pool replication results. In a separate re-optimization pilot on 20 jailbroken JailbreakBench prompts, the full perturbation reaches ASR $\SuffAdvFull$ ($19/20$). Restricting optimization to the tested least-important band $[20,29]$, top band $[70,79]$, or their union gives ASRs $\SuffBandLow$, $\SuffBandTop$, and $\SuffBandUnion$. These tested supports do not cover all possible subsets. Random supports covering about half the STFT bins reach mean ASR $0.86$ in the same pilot, showing that the full band set is not required in these trials. The equal-Hz and matched-energy scattered-removal checks use the available selected subsets and should be read with the same selection caveat. Raw depth correlations rise in the three existing 50-prompt samples ($\rho_0=+0.69,+0.87,+0.72$ to $\rho_{30}=+0.88,+0.98,+0.93$), but none rejects the phase-randomized late-slope null ($p=\PJbbP$, $\PMiP$, $\PSrP$; primary $p=\DepthSurrP$).

\vspace{\ParaVspace}
\textbf{Qualifications.} The additional corpora use a different renderer, so absolute ASRs are not directly comparable with AdvBench. Only the JailbreakBench $K{=}8$ mask sweep uses the full jailbroken pool; MaliciousInstruct and StrongREJECT, and the auxiliary analyses described above, use necessity-selected 50-prompt samples. Those selected samples cannot independently confirm the ranking used to select them. The three depth curves are also based on these existing 50-prompt samples and none rejects the phase-randomized null ($p=\PJbbP$, $\PMiP$, $\PSrP$; primary $p=\DepthSurrP$). We treat the cross-corpus evidence as partial and descriptive, not as full replication.

\end{document}